\documentclass[a4paper,11pt]{article}
\pdfoutput=1 

\usepackage{enumitem}
\usepackage{adjustbox}
\usepackage[T1]{fontenc} 
\usepackage{enumitem}  
\usepackage{empheq}
\usepackage{dsfont}
\usepackage[mathscr]{euscript}
\usepackage{dcolumn}
\usepackage{bm}
\usepackage{bbm}
\usepackage{slashed}
\usepackage{wrapfig}
\usepackage{mathtools}
\usepackage{tikz}

\usepackage{jheppub} 
\usepackage{amsfonts,cases}
\usepackage{amsthm}

\newcommand{\vol}[1]{\text{vol}(#1)~}
\newcommand{\beq}{\begin{equation}}
\newcommand{\eeq}{\end{equation}}
\newcommand{\bea}{\begin{eqnarray}}
\newcommand{\eea}{\end{eqnarray}}
\newcommand{\eg}{{\it e.g.,}\ }

\def\Tr{{\rm Tr}}

\newcommand{\pd}{\partial}

\newcommand{\eps}{\varepsilon}

\newcommand\eq[1]{eq.~(\ref{eq:#1})}

\newcommand{\sn}[1]{section~\ref{sec:#1}}

\newcommand{\app}[1]{appendix~\ref{app:#1}}

\newcommand\mf[1]{{\mathfrak{#1}}}

\newcommand\CF{{\mc{F}}}

\newcommand\CI{{\mc{I}}}

\newcommand\CN{{\mc{N}}}
\newcommand\CO{{\mc{O}}}

\newcommand\CQ{{\mc{Q}}}

\newcommand\II{{\rm I\hspace{-0.02cm} I}}
\newcommand\oII{\mathring{\II}}

\definecolor{cardinal}{rgb}{0.6,0,0}
\definecolor{darkgreen}{rgb}{0,0.5,0}
\definecolor{darkblue}{rgb}{0.2, 0, 0.8}
\definecolor{golden}{rgb}{0.92, 0.7, 0}

\tikzset{flavour/.style={draw=none,minimum size=0.3mm,fill=white, regular polygon,regular polygon sides=4,draw}}
\tikzset{gaugeBig/.style={inner sep=1mm,draw=none,fill=white,minimum size=2mm,circle, draw}}
\tikzset{bd/.style={circle, draw=black, inner sep=0pt, fill=black, minimum size=2mm}}
\tikzset{wd/.style={circle, draw=black, inner sep=0pt, fill=white, minimum size=2mm}}
\tikzset{Dynkin/.style={circle, draw=black, inner sep=0pt, fill=white, minimum size=2mm}}
\tikzstyle{ligne}=[draw, very thick] 
\tikzstyle{gridline}=[draw, gray] 
\tikzset{gauge/.style={circle, draw,inner sep=2.5pt}}
\tikzset{gaugeo/.style={circle, draw,inner sep=2.5pt,fill=orange}}
\tikzset{gaugec/.style={circle, draw,inner sep=2.5pt,fill=cyan}}
\tikzset{gauger/.style={circle, draw,inner sep=2.5pt,fill=red}}
\tikzset{gaugeb/.style={circle, draw,inner sep=2.5pt,fill=blue}}
\tikzset{gaugeg/.style={circle, draw,inner sep=2.5pt,fill=green}}
\tikzset{gaugem/.style={circle, draw,inner sep=2.5pt,fill=magenta}}
\tikzset{hasse/.style={circle, fill,inner sep=2pt}}
\tikzset{shrinky/.style={circle, fill,inner sep=1pt}}
\tikzset{sized/.style={circle, draw, inner sep=1.5pt}}
\tikzset{seven/.style={circle, draw,inner sep=3pt}}

\tikzset{dotto/.style={circle, orange, draw,inner sep=1.5pt,fill=orange}}
\tikzset{dottp/.style={circle, purple, draw,inner sep=1.5pt,fill=purple}}
\tikzset{dottc/.style={circle, cyan, draw,inner sep=1.5pt,fill=cyan}}
\tikzset{dottr/.style={circle, red, draw,inner sep=1.5pt,fill=red}}
\tikzset{dottb/.style={circle, blue, draw,inner sep=1.5pt,fill=blue}}
\tikzset{dottg/.style={circle, green, draw,inner sep=1.5pt,fill=green}}
\tikzset{dottm/.style={circle, magenta, draw,inner sep=1.5pt,fill=magenta}}

\DeclareMathOperator{\SU}{SU}

\usepackage{epsfig}
\usepackage{amsmath}
\usepackage{amssymb}
\usepackage{amsfonts}
\usepackage{amsxtra}
\usepackage{amsthm}
\usepackage{mathrsfs}
\usepackage{makeidx}
\usepackage{graphics}
\usepackage{dsfont}
\usepackage{mathtools}
\usepackage{graphicx}
\usepackage{placeins}
\usepackage{bm}
\usepackage[capitalise]{cleveref}
\usepackage{empheq}
\usepackage{colortbl}
\usepackage{xcolor}
\usepackage{titlesec}
\usepackage{longtable}
\usepackage{hhline}
\usepackage{float}
\usepackage{color}
\usepackage{tikz}
\usepackage{xfrac}
\usepackage{footnote}
\usepackage{rotating}
\usepackage{lscape}
\usepackage{makecell}
\usepackage{environ}
\usepackage{tabularx}
\usepackage{subfiles}
\usepackage{ytableau}
\usepackage{tikz-3dplot}
\usepackage{slashed}
\usepackage{pifont}
\usepackage{multirow}
\usepackage{mdframed}
\usepackage{bbm}
\usepackage[normalem]{ulem}
\usepackage{arydshln}
\usepackage{enumitem} 
\usepackage{fancybox}

\usetikzlibrary{positioning,trees,decorations.pathmorphing,decorations.markings,decorations.pathreplacing,calc,shapes,patterns,arrows,chains,arrows.meta,fit,fadings,decorations.markings,graphs,graphs.standard,quotes,plotmarks,calligraphy,decorations.pathreplacing}

\usetikzlibrary{positioning}
\usetikzlibrary{chains}
\usetikzlibrary{arrows, arrows.meta ,fit,decorations.pathreplacing}
\tikzstyle{every picture}+=[remember picture]
\tikzstyle{na} = [baseline]
\tikzstyle{ligne}=[draw, thick]

\usetikzlibrary{arrows, decorations.markings, calc, fadings, decorations.pathreplacing, patterns, decorations.pathmorphing, positioning}
\tikzset{>={Latex[width=1.5mm,length=1.5mm]}}
\tikzset{bd/.style={circle, draw=black, inner sep=0pt, fill=black, minimum size=1.2mm}}
\tikzset{bld/.style={circle, draw=blue, inner sep=0pt, fill=blue, minimum size=1.2mm}}
\tikzset{wd/.style={circle, draw=black, inner sep=0pt, fill=white, minimum size=1.2mm}}
\tikzset{rd/.style={circle, draw=red, inner sep=0pt, fill=red, minimum size=.9mm}}
\tikzset{wrd/.style={circle, draw=red, inner sep=0pt, fill=white, minimum size=.9mm}}
\usetikzlibrary{graphs,graphs.standard,quotes}

\numberwithin{equation}{section}
\newcommand{\bes}[1]{\begin{equation} \begin{split} #1\end{split} \end{equation}}

\newcommand{\be}{\begin{equation}} \newcommand{\ee}{\end{equation}}

\def\tilde{\widetilde}

\def\hat{\widehat}

\def\bar{\overline}

\def\rt2{\sqrt{2}}

\def\det{\mathop{\rm det}}

\def\Tr{\mathop{\rm Tr}}

\def\CF{{\cal F}}

\def\CI{{\cal I}}

\def\CN{{\cal N}}
\def\CO{{\cal O}}

\def\CQ{{\cal Q}}

\def\1{{\ds 1}}

\newcommand{\cQ}{\mathcal{Q}}

\def\SU{\mathrm{SU}}

\def\su{\mathfrak{su}}

\def\so{\mathfrak{so}}

\def\repa{\raise4pt\hbox{$\square$}\mkern-14mu\raise-4pt\hbox{$\square$}}
\def\repab{\overline{\raise4pt\hbox{$\square$}\mkern-14mu\raise-4pt\hbox{$\square$}\mkern-1mu}}

\newcommand{\ba}{\begin{array}}
\newcommand{\ea}{\end{array}}
\newcommand{\bi}{\begin{itemize}}
\newcommand{\ei}{\end{itemize}}

\def\bea#1\eea{\allowdisplaybreaks \begin{align}#1\end{align}}
 \newcommand{\ben}{\begin{enumerate}}
\newcommand{\een}{\end{enumerate}}
\newcommand{\bean}{\begin{eqnarray*}}
\newcommand{\eean}{\end{eqnarray*}}

\definecolor{light-gray}{gray}{0.5}
\definecolor{new-green}{rgb}{0,0.7,0.3}

\newcommand{\blue}{\color{blue}}

\newcommand{\red}{\color{red}}

\tikzset{snake it/.style={decorate, decoration={snake, amplitude=.4mm, segment length=2mm,
                       post length=0mm,pre length=0mm}}}

\makeatletter
\gdef\@fpheader{}
\makeatother

\begin{document}

\title{Superconformal anomalies for string defects\\ in six-dimensional $\CN = (1,0)$ SCFTs}

\author[1,2]{Fabio Apruzzi,}
\author[3,4]{Noppadol Mekareeya,}
\author[3]{Brandon Robinson,}
\author[3,5]{Alessandro Tomasiello}

    \affiliation[1]{Dipartimento di Fisica e Astronomia “Galileo Galilei”, Università di Padova,
    Via Marzolo 8, 35131 Padova, Italy}
    \affiliation[2]{INFN, Sezione di Padova Via Marzolo 8, 35131 Padova, Italy}
    \affiliation[3]{INFN, Sezione di Milano-Bicocca, Piazza della Scienza 3, I-20126 Milano, Italy}
    \affiliation[4]{Department of Physics, Faculty of Science, Chulalongkorn University, Phayathai Road, Pathumwan, Bangkok 10330, Thailand}
    \affiliation[5]{Dipartimento di Matematica, Università di Milano–Bicocca, Via Cozzi 55, 20126 Milano, Italy}

\emailAdd{fabio.apruzzi@pd.infn.it}
\emailAdd{n.mekareeya@gmail.com}
\emailAdd{brandon.robinson@mib.infn.it}
\emailAdd{alessandro.tomasiello@unimib.it}

\abstract{
We study the anomalies of two-dimensional BPS defects in six-dimensional $\CN=(1,0)$ superconformal field theories.  Using a holographic description of these defects furnished by probe D4-branes in AdS${}_7$ solutions of ten-dimensional type IIA supergravity, we compute the two independent defect Weyl anomalies from the on-shell action for a spherical defect and defect sphere entanglement entropy.  We find agreement between the holographic prediction for the defect A-type anomaly coming from the defect sphere free energy and the leading large $N$ contribution to the defect `t Hooft anomaly found using anomaly inflow.  We also find agreement between the holographic computation of the expectation value of a surface operator wrapping a torus and the supersymmetric localization computation for a circular Wilson loop in $\CN=1$ super Yang-Mills theory on $\mathds{S}^5$. Lastly, we holographically compute the defect gravitational anomaly from the Wess--Zumino action of the probe D4-brane, which provides a subleading large $N$ correction to the defect A-type anomaly.
}
\preprint{}
\maketitle

\section{Introduction}

Among the powerful holographic tools available to study gauge theories at strong coupling, probe brane holography provides a controlled framework with which to systematically introduce heavy charged operators \cite{Karch:2002sh} or submanifold localized degrees of freedom \cite{DeWolfe:2001pq, Constable:2002xt} in a regime where the number of operators holographically sourced by the probe brane remains parametrically small compared to the number of color degrees of freedom. By carefully constructing the brane embedding to preserve supersymmetry on the intersection of the curved conformal boundary of the ambient AdS space and the brane worldvolume, probe brane holography has passed a number of precision tests through comparison to results obtained by supersymmetric localization \cite{Karch:2015vra, Karch:2015kfa, Robinson:2017sup}.  On the gravity side, taking the probe limit of a brane embedding includes taking the brane tension to be parametrically large, which to leading order freezes out the probe brane's gravitational degrees of freedom and suppresses backreaction onto the background geometry.  This limit greatly simplifies the computation of key physical quantities that characterize probe brane degrees of freedom such as the brane's on-shell action and contributions to Entanglement Entropy (EE), both of which will be the primary focus of our holographic computations.

In the study of superconformal field theories (SCFTs), the $d=6$ case has a distinguished role. It is the highest dimension in which SCFTs exist; moreover, six-dimensional (6d) theories spawn a large variety of theories in lower dimensions. The most famous example is the ${\mathcal N}=(2,0)$ theory on M5-brane stacks \cite{Witten:1995zh}, but a lot more examples exist with ${\mathcal N}=(1,0)$, engineered by orbifolds, M-theory walls, F-theory, IIA brane intersections \cite{Intriligator:1997dh,Ganor:1996mu,Heckman:2013pva,Hanany:1997gh}. The latter also have holographic duals, the AdS$_7\times M_3$ solutions in massive IIA \cite{Apruzzi:2013yva,Apruzzi:2015wna,Cremonesi:2015bld}. 

Codimension-two defects in 6d SCFTs have played a role in their compactifications down to four dimensions, appearing as punctures on an internal Riemann surface; see for example \cite{Gaiotto:2009we,Gaiotto:2009gz,Agarwal:2014rua,Razamat:2022gpm}. On the other hand, codimension-four defects have been studied much less. In   ${\mathcal N}=(2,0)$ theories they are provided by M2-branes \cite{Alday:2009fs,Drukker:2010jp,Mori:2014tca}. In this paper we initiate the study of codimension-four defects in the much more numerous ${\mathcal N}=(1,0)$ theories.

The generic form for the Weyl anomaly of a (super)conformal defect supported on a two-dimensional submanifold, $\Sigma$, embedded a $d$-dimensional ambient space is\footnote{Since our focus is on co-dimension four defects, there are only parity even defect anomalies \cite{Jensen:2018rxu}.}
\begin{align}\label{eq:defect-weyl-anomaly}
    T^\mu{}_\mu|_\Sigma = \frac{1}{24\pi}(a_\Sigma \overline{E}_2+ d_1\oII^2 + d_2 P[W]_\Sigma)
\end{align}
where $\overline{E}_2$ is the two-dimensional Euler density built from the intrinsic metric of the defect submanifold, $\oII^2$ is the square of the trace-free second fundamental form for the embedding, and $P[W]_\Sigma$ is the trace of the pullback of the ambient Weyl tensor to the defect. The A-type \cite{Deser:1993yx} defect Weyl anomaly $a_\Sigma$ has been shown to obey a weak c-theorem \cite{Jensen:2015swa, Wang:2020xkc} and, with a sufficient amount of supersymmetry preserved on the defect, to be related to defect R- and gravitational anomalies \cite{Wang:2020xkc}.  The B-type anomalies $d_1$ and $d_2$ are strictly non-negative in reflection positive theories \cite{Billo:2016cpy, Jensen:2018rxu}, but do not obey any known c-theorem.  However, relevant to the cases we are interested, it is known that $d_1=d_2$ for two-dimensional conformal defect preserving at least $\CN=(2,0)$ defect supersymmetry \cite{Bianchi:2019sxz,Drukker:2020atp}, and so in order to characterize BPS string defects in six-dimensional $\CN=(1,0)$ SCFTs, we only need to compute $a_\Sigma$ and $d_2$.  Crucial for our analysis below, these defect anomalies are known to control two quantities that can be easily computed using probe brane holography: the log divergent part of defect sphere free energy, which is uniquely determined by $a_\Sigma$, and the universal part of the defect sphere EE, which is fixed by a linear combination of $a_\Sigma$ and $d_2$ \cite{Kobayashi:2018lil,Jensen:2018rxu}.

In \sn{ads7}, we briefly review AdS$_7$ solutions to type IIA SUGRA found in \cite{Apruzzi:2013yva,Apruzzi:2015wna,Cremonesi:2015bld} and their holographic dual description in terms of six-dimensional $\CN=(1,0)$ SCFT at large $N$. 

In \sn{probes}, we construct solutions for embedding probe D4-brane into a generic AdS$_7\times M_3$ background with non-trivial Romans mass.\footnote{Some fully back-reacted solutions for codimension-four defects exist \cite{DHoker:2008lup,Dibitetto:2017klx,Lozano:2019ywa,Lozano:2022ouq}, and it would be interesting to compare our approach to those results; the AdS$_3\times \mathds{S}^3$ solutions appear particularly relevant.}  Further in this section, we carry out four holographic computations that characterize the two-dimensional defect in the field theory.  First, we compute the defect contribution to the EE of a spherical region.  Second, we compute the $a_\Sigma$ anomaly from the free energy of a spherical defect and use the defect EE result to obtain the B-type anomalies. In terms of the number of probe D4-branes $q_i$ and partitions of D6-branes $r_i$ they can be succinctly written to leading order in large $N$ as
\begin{align}
    a_\Sigma = 24(q,r),\qquad d_1=d_2= 32(q,r)~
\end{align}
with scalar product taken with respect to the inverse Cartan matrix of $A_{N-1}$. Third, we holographically compute the expectation value of the string defect operator $W$ on $\mathds{S}^1_{\beta/4\pi}\times\mathds{S}^1$, which gives
\begin{align}
    \left< W \right> = \exp\left[\beta (q,r)\right].
\end{align}
Lastly, we compute the defect gravitational anomaly from the Wess--Zumino action, which is shown to be subleading at large $N$ and contributes to the $a_\Sigma$ anomaly
\begin{align}
    a_\Sigma = 24(q,r) + \frac{1}{2}\sum_{i}q_i r_i\,.
\end{align}

In \sn{ft-anomalies}, we compare the probe brane results to anomalies computed via other means. In \sn{loc}, we compare the expectation value of the torus Wilson surface to the dimensionally reduced circular Wilson loop obtained purely in field theory using supersymmetric localization of $\CN=1$ SYM theory on $\mathds{S}^5$.  Despite the holographic computation only computing the value at large $N$ and the supersymmetric localization result being exact, we find a precise match between the two expressions. Later, in \sn{inflow}, we consider the type IIB description of string defects in six-dimensional SCFTs. By the help of these string constructions we study the 6d anomaly polynomial inflow onto the defect. In particular, we compute the defect R-anomaly and successfully match the leading large $N$ result for the $a_\Sigma$ anomaly. As a byproduct we developed the inflow mechanism for (string) defects provided by probe branes in IIB backgrounds engineering 6d SCFTs. This approach could be potentially generalized to other type (non-string) of defects.

In \app{conformal}, we explain the coordinate transformations used in \sn{embedding}.  In \app{2dquiver}, we consider a purely two-dimensional description of the D2--D4--D6 brane system in terms of a quiver gauge theory.  We demonstrate that the quiver description correctly captures the subleading large $N$ contributions (only) to the defect anomaly, i.e. the gravitational anomaly. Differently from the BPS strings, which become tensionless at the fixed point, the quiver which includes the defect does not capture the full defect degrees of freedom. We argue that the quiver only provides the intrinsic degrees of freedom of the defect. However it does not capture the bulk-defect ones, which are instead computed via anomaly inflow.

\section{\texorpdfstring{AdS$_7$/CFT$_6$}{AdS7/CFT6}}\label{sec:ads7}

The starting point is the asymptotically AdS$_7 \times M_3$ background solutions to type IIA SUGRA constructed in \cite{Apruzzi:2013yva,Apruzzi:2015wna,Cremonesi:2015bld}
\begin{align} \label{eq:ads7}
    \begin{split}
       \frac{1}{\pi\sqrt{2}}d s^2 &= 8 \sqrt{-\frac{\alpha}{\alpha^{\prime\prime}}}ds_7^2 + \sqrt{-\frac{\alpha^{\prime\prime}}{\alpha}}\left(dz^2 +\frac{\alpha^2}{{\alpha^\prime}^2 - 2\alpha\alpha^{\prime\prime}}d\Omega_2^2\right)\,,\\
       e^\phi &= 3^4(\sqrt{2}\pi)^\frac{5}{2}\frac{(-\alpha/\alpha^{\prime\prime})^{\frac{3}{4}}}{\sqrt{{\alpha^\prime}^2 -2\alpha\alpha^{\prime\prime}}}\,,\\
       B & = \pi\left(-z + \frac{\alpha\alpha^\prime}{{\alpha^\prime}^2 - 2\alpha\alpha^{\prime\prime}}\right)\text{vol}_{\mathds{S}^2}\,,\\
       F_2& = \left(\frac{\alpha^{\prime\prime}}{162\pi^2}+\frac{\pi F_0 \alpha\alpha^\prime}{{\alpha^\prime}^2 - 2 \alpha\alpha^{\prime\prime}}\right)\text{vol}_{\mathds{S}^2}\,.
    \end{split}
\end{align}
Here $d s_7^2$ is the line element on AdS$_7$, $d\Omega^2_2$ is the line element on the unit 2-sphere $\mathds{S}^2$,
prime denotes $\partial_z$, and $\alpha$ is a $C^1$ piecewise-cubic function on the interval $z\in[0,N]$ that satisfies $\alpha>0$, $\alpha''<0$,
\begin{equation}\label{eq:a3}
    \alpha'''(z) = -162 \pi^3 F_0
\end{equation}
almost everywhere, and appropriate boundary conditions at the endpoints of the interval. The simplest possibility is $\alpha=\alpha''=0$, which ensures smoothness, but at either endpoint it is also possible to introduce D6-branes (with $\alpha=0$), O8-planes ($\alpha'=0$), O6-planes ($\alpha''=0$). 

NSNS flux quantization implies $N\in {\mathbb Z}$, while for RR it gives
\begin{equation}\label{eq:fq}
     -\frac{\alpha''(i)}{81 \pi^2}\in {\mathbb Z}\,,\qquad \forall i \in {\mathbb Z}\,.
\end{equation}
$F_0=\frac{n_0}{2\pi}$, $n_0\in {\mathbb Z}$ is locally constant, but can jump across D8-branes; these are restricted to lie at integer values $ \mu$ of $z$, which are in turn identified with their D6-brane charge. A cartoon of the internal geometry is shown in Fig.~\ref{fig:ads7-D4D2}.

The curvature and string coupling of any given solution can be rescaled and made arbitrarily small by taking 
\begin{equation}\label{eq:hol}
    N\gg 1\, \qquad \mu/N \ \text{finite}\,.
\end{equation}
This is the regime relevant for holography.

The simplest solution occurs when the Romans mass $F_0=0$. \eqref{eq:a3} implies that $\alpha''$ is constant, and \eqref{eq:fq} implies
\begin{equation}\label{eq:massless}
    \alpha= \frac{81}2 \pi^2 {k} z (N-z)\,.
\end{equation}
${k}\in {\mathbb Z}$ is interpreted as the $F_2$ flux integer on the $\mathds{S}^2$. There are ${k}$ D6-branes at $z=0$ and $-{k}$ anti D6-branes at $z=N$. The resulting IIA solution is the dimensional reduction of the ${\rm AdS}_7\times\mathds{S}^4/\mathbb{Z}_{k}$ solution of eleven-dimensional SUGRA. 

When $F_0\neq 0$, $\alpha''$ is piecewise linear; it can be parameterized as 
\begin{equation}\label{eq:a''-ri}
    -\frac{\alpha''(z)}{81 \pi^2} = r_i + (r_{i+1}-r_i) (z-i)
\end{equation}
on each interval $z\in [i,i+1]$. By \eq{fq}, the $r_i\in {\mathbb Z}$.

These AdS$_7$ solutions were conjectured to arise from the near-horizon limit of NS5--D6--D8 brane diagrams, and this led to the identification of the dual SCFT$_6$ \cite{Gaiotto:2014lca}. We now give a quick review focused on what we need in this paper; see \cite[Sec.~2.1]{Cremonesi:2015bld} for more details.
   
The brane diagram can be described as follows. There are $N$ NS5-branes along directions $012345$, separated along direction $5$. Between the $(i-1)^{\rm th}$ and $i^{\rm th}$ NS5-brane,
$r_i$ D6-branes are suspended, along directions $0123456$; crossing the latter, there are also $f_i=2r_i - r_{i+1}-r_{i-1}$ D8-branes along $012345789$. This brane diagram is depicted schematically in the black part of Fig.~\ref{fig:brane-diagram-D4D2}.

The six-dimensional field theory engineered by this diagram is a chain of $N-1$ vector multiplets with gauge groups ${\rm U}(r_i)$; there are hypermultiplets in each of the bifundamental representations $\overline{\bf r_i}\otimes {\bf r_{i+1}}$, and $f_i$ in the fundamental ${\bf r_i}$. There are also $N-1$ tensor multiplets. 

\eqref{eq:a''-ri} tells us that the function $r(z)\equiv -\alpha''/81\pi^2$ gives the ranks $r_i$ when evaluated at $z=i$; conversely we can write 
\begin{equation}\label{eq:a-r}
    \alpha(z)= - 81 \pi^2 \frac1{\partial_z^2} r(z) \,.
\end{equation}
The integration constants in the double integral should be understood as being fixed by the boundary conditions. As an example, when $F=0$ all the ranks are equal, $r_i=k$, and it is easy to see that \eq{a-r} reproduces \eq{massless}.

An important piece of evidence for the AdS$_7$/CFT$_6$ conjecture we just reviewed came from a holographic match of the $a$ anomaly \cite{Cremonesi:2015bld, Apruzzi:2017nck}. At leading order in the limit \eq{hol}, the gravitational result reads
\begin{equation}\label{eq:a6d-hol}
    a_{\rm 6d, hol}= \frac{64}{189\pi^2}\int r \alpha \,dz = -\frac{192}7 \int r\frac1{\partial_z^2} r \,d z \,.
\end{equation}
In the second step we have used \eq{a-r}. On the other hand, the leading term in the field theory computation is given by a Green--Schwarz term and reads
\begin{equation}\label{eq:a6d}
    a_{\rm 6d} = \frac{192}7 (C^{-1})^{ij} r_i r_j\,,
\end{equation}
where $C_{ij}$ is the Cartan matrix of $A_{N-1}$. $C_{ij}$ is a discrete analogue of minus a double derivative: indeed
\begin{equation}\label{eq:cartan-der}
    C_{ij}r_j = -r_{i+1}+2 r_i -r_{i-1} \sim - (\partial^2 r)_i
\end{equation}
So its inverse is a discrete analogue of a double integral. Hence \eq{a6d} matches \eq{a6d-hol} when $N\gg 1$. 

To summarize the important lesson for our present purposes, in the holographic limit we have 
\begin{equation}\label{eq:a-Cr}
    \alpha^i = 81\pi^2 (C^{-1})^{ij} r_j\,.
\end{equation}
As an example, we can again look at the $F_0=0$ case, where $r_i=k$ and $(C^{-1})^{ij}r_j= \frac12 i(N-i)$, in agreement with \eq{massless}.

\section{Probe brane holography}\label{sec:probes}

In this section, we construct embeddings for probe D4-branes in the asymptotically AdS$_7\times M_3$ geometries reviewed in \sn{ads7} that are dual to BPS codimension-4 defects in six-dimensional $\mathcal{N} =(1,0)$ SCFTs. Using these solutions for embedded AdS$_3$ branes, we characterize these defects by computing their independent defect Weyl anomalies using probe brane holography.  First, we holographically compute the defect contribution to the EE of a spherical region in the dual field theory; this furnishes for us a specific linear combination of the two independent defect Weyl anomalies. Second, we compute the free energy of the defect wrapping $\mathds{S}^2\hookrightarrow\mathbb{R}^6$, which gives us the A-type defect anomaly $a_\Sigma$.  The value of $a_\Sigma$ matches the predicted value computed from in-flow, and so, bolsters the standing of the holographic prediction for the B-type anomaly $d_2$.  Lastly, compute the defect supersymmetry Casimir Energy (SCE), which is obtained holographically from the on-shell action of a probe brane wrapping $\mathds{S}^1\times\mathds{S}^1\hookrightarrow\mathds{S}^1\times\mathds{S}^5$.

\subsection{Probe D4-brane embedding}\label{sec:embedding}

Throughout this section, we will work in Euclidean signature. The probe brane action is 
\begin{align}
    S_{\rm probe} = T_4 \int_{M_5}d^5x e^{-\phi}\sqrt{\det P[g + B] + \CF} - T_4\int_{M_5} e^{\CF}\wedge P[C]\,,
\end{align}
where 
\begin{equation}
    T_4 = \frac1{16\pi^4 l_s^5}\,;
\end{equation}
$M_5$ is the D4-brane worldvolume with coordinates $x^a$, $P$ denotes pull-back onto $M_5$, and the worldvolume gauge field $\CF= B + 2\pi l_s f$.   Throughout, we will work with convention $l_s =1$.

We will consider D4-brane probes of the factorized form
\begin{equation}\label{eq:fac}
    \tilde M_3\times \mathds{S}^2 \subset {\rm AdS}_7 \times M_3\,,
\end{equation}
where $\tilde M_3\subset {\rm AdS}_7$ and the $\mathds{S}^2$ is the sphere appearing in \eq{ads7}, whose $z$ position we will determine shortly. Our D4-branes will carry non-zero D2-brane charge $j\in {\mathbb Z}$: 
\begin{equation}\label{eq:f-ws}
    f = \frac j2  \text{vol}_{\mathds{S}^2}\,.
\end{equation}

The position $z$ of the D4 can be fixed by extremizing the action. This is implied by the BPS condition, which in general reads $\Gamma_\parallel \epsilon_2= \epsilon_1$ for a D-brane, with $\Gamma_\parallel$ the product of the parallel gamma matrices and $\epsilon_a$ the supersymmetry parameters preserved by the solution. In our case, the $\epsilon_a$ are factorized in terms of Killing spinors $\zeta$ in AdS$_7$ and internal spinors $\chi_a$ \cite[(A.4)]{Apruzzi:2013yva}. $\Gamma_\parallel$ is also factorized because of our assumption \eqref{eq:fac}.

The internal part $\gamma_\parallel \chi_1= \chi_2$ of the BPS computation can be carried out efficiently in terms of calibrations. It is in fact identical to that for a D8/D6 bound state in \cite[Sec.~4.8]{Apruzzi:2013yva}, so we will be brief. (This happens because intersecting or overlapping D-branes preserve supersymmetry if the number of Neumann--Dirichlet directions is a multiple of 4.) The relevant condition is\footnote{In the language of \cite{Apruzzi:2013yva}, the calibration is ${\rm Im} \psi^1_+$; alternatively, one can impose that the forms $({\rm Im} \psi^2_+, {\rm Re}  \psi^2_+, {\rm Re}  \psi^1_+)$ pull back to zero.}  
\begin{align}
    {\cal F} = \pi \frac{\alpha \alpha^\prime}{{\alpha^\prime}^2 - 2 \alpha \alpha^{\prime\prime}}\text{vol}_{\mathds{S}^2}.
\end{align}
Using \eqref{eq:ads7}, \eqref{eq:f-ws}, this implies
\begin{equation}
    z=j\,.
\end{equation}
As anticipated, this coincides with the BPS position for a D8-brane with D6 charge equal to $j$, which was already noticed below \eqref{eq:fq}.

\begin{figure}[t]
\centering	
\includegraphics[width=8cm]{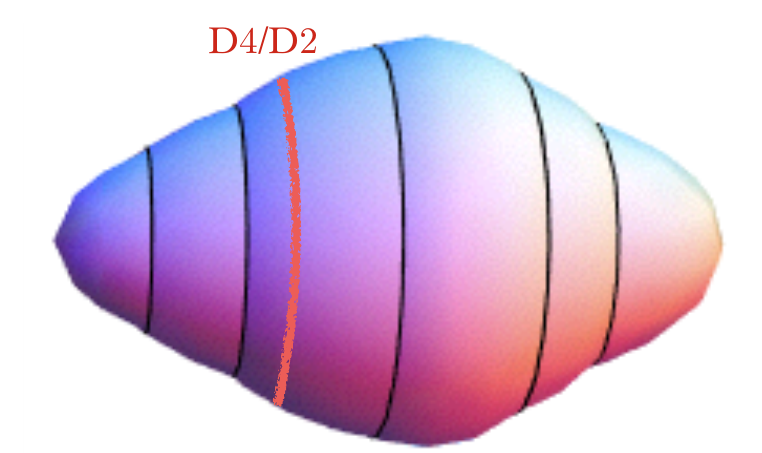}
\caption{A cartoon of the internal three-dimensional space of the AdS$_7$ solution. The black creases represent the D8-branes, where the metric is continuous but not differentiable. The red locus represents a D4--D2 bound state probe, to be discussed in the next section.}
\label{fig:ads7-D4D2}
\end{figure}

\begin{figure}[ht]
\centering	
\includegraphics[width=10cm]{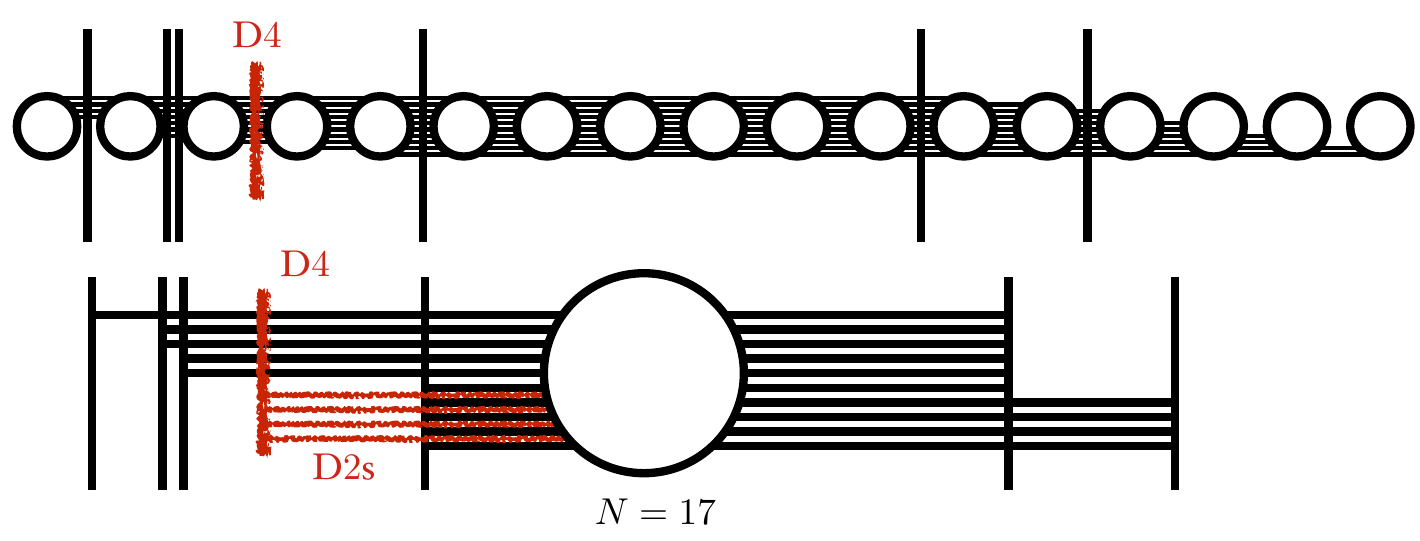}
\caption{\small The black part is a schematic representation of the NS5--D6--D8 brane diagram, taken from \cite{Cremonesi:2015bld}. The nodes, horizontal lines and vertical lines represent NS5-, D6- and D8-branes respectively. The upper part of the diagram, where the NS5 are separated, is more useful to read the field theory. The red vertical lines here represent additional D4-branes, which engineer line defects in the field theory, to be discussed in the next section.
The lower part, where the NS5-branes coincide, corresponds to the conformal phase of the field theory, and resembles more directly the gravity solution. The brane-creation effect changes the number of D6-branes, and leads to the creation of D2-branes, represented by red horizontal lines. In the near-horizon limit, each D8 with D6s ending on it becomes a D8--D6 bound state, and the D4 with D2s becomes a D4--D2 bound state.}
\label{fig:brane-diagram-D4D2}
\end{figure}

The case of a single D2 has to be treated separately; again this works in the same way as a D6 extended along all of AdS$_7$. The calibration analysis this time gives $\alpha \alpha''=0$, which is only true at the extrema of the interval, $z=0$ and $z=N$. The resulting probe D4--D2 bound state wraps an internal $\mathds{S}^2$ at fixed $z=j$, which is schematically visualized in Figures \ref{fig:ads7-D4D2} and \ref{fig:brane-diagram-D4D2}.

The probe brane action then takes the form 
\begin{align}
    \label{eq:D4-probe-action-OS}
    S_{\rm probe} &= T_4 \int_{M_5}d^5x  e^{3A-\phi}f_1^2f_2\sqrt{g_{\theta\theta}g_{\phi\phi}+ {\cal F}_{\theta\phi}^2}\big|_{z=j}\\
   &\nonumber = 
   \frac{64\pi}{81}T_4\vol{\tilde M_3}\sqrt{\alpha'^2-\alpha \alpha''}
   \left(\left(\sqrt{-\frac{\alpha''}\alpha}\frac{\sqrt 2 \alpha^2}{\alpha'^2 - 2 \alpha \alpha''}\right)^2 + \left(\frac{\alpha \alpha'}{\alpha'^2 - 2 \alpha \alpha''}\right)^2\right)^{1/2}\\
   &\nonumber
   =\frac{64 \pi }{81}T_4\vol{\tilde M_3}\alpha^j \,,
\end{align}
where we introduced $\alpha^j\equiv \alpha(j)$. In the second line, we have used \eqref{eq:ads7}. Since the D2 charge carried by the probe D4 brane eliminated the explicit $z$ dependence, the action completely factorized into a product of an integral over the volume form of the $\tilde M_3$ and an integral over the internal $\mathds{S}^2$ geometry threaded by two-form flux.

In what follows we will consider a more general situation where there is a collection of $q_j$ D4-branes with D2-brane equal to $j$. This will turn \eqref{eq:D4-probe-action-OS} into $\frac{64 \pi }{81}T_4\vol{\tilde M_3} q_j \alpha^j$. We will work in the probe approximation: this requires $q_j e^{\phi(j)}\ll 1$, which by \eqref{eq:ads7} is implied by $q_j \ll \sqrt{N}$.

It is clear from the form of the on-shell action in \eq{D4-probe-action-OS} that we will need to regulate divergences coming from $\vol{\tilde M_3}$. To that end, we will use the  holographic renormalization scheme for probe branes developed in \cite{Karch:2005ms}. As a brief review, this is done by introducing a large radial cutoff $\Lambda$ in the background AdS$_7$ geometry and computing the covariant counterterms on the intersection of the radial hypersurface and the worldvolume of the probe brane.  We denote coordinates on this intersection by $y^a$ and the induced metric by $\gamma_\Lambda$.  Since, we have no worldvolume scalars turned on in our embedding, the only counterterm that we need to compute is the volume of the intersection
\begin{align}\label{eq:counterterm-action}
    S_{\rm CT} =  -\frac{2\pi}{9}T_4\sqrt{2r_j\alpha^j}  \int d^2 y \sqrt{\gamma_\Lambda}\,,
\end{align}
where the coefficient is fixed to cancel polynomial divergences in $\Lambda$ appearing in $\vol{\tilde M_3}$.  In the subsections below, we denote the holographically renormalized probe brane action by
\begin{align}\label{eq:renormalized-os}
    S_{\rm ren} = S_{\rm probe} + S_{\rm CT}\,.
\end{align}

\subsection{Spherical entanglement entropy}

Here we compute the contribution of the probe string defect to the spherical entanglement entropy (EE).  For simplicity, we will consider a flat embedding of a two-dimensional defect supported on $\mathbb{R}^{1,1}$ in $\mathbb{R}^{5,1}$, which will be taken as the conformal boundary of the AdS$_7$ part of the background solution.  Since we are working entirely within the probe limit, we do not need to compute the backreaction and can compute the EE by using \cite{Casini:2011kv}.

First, we consider the AdS$_7$ part of the background geometry in flat slicing 
\begin{align}\label{eq:flat-sliced-ads7}
    ds^2_7 = \frac{1}{u^2}(du^2-dt^2+ dx_1^2+d\xi^2 + \xi^2 d\Omega_3^2)\,,
\end{align}
with the probe D4-brane wrapping $\{t,x_1,u\}$ and $d\Omega_3^2$ the line element on the unit $\mathds{S}^3$.  Following \cite{Casini:2011kv, Jensen:2013lxa}, we map 
\begin{align}
    \begin{split}\label{eq:CHM-map}
        t &= \frac{R\sqrt{\rho^2-1}\sinh\tau}{\rho\cosh\upsilon +\sqrt{\rho^2-1}\cosh\tau},\hspace{2cm} u = \frac{R}{\rho\cosh\upsilon +\sqrt{\rho^2-1}\cosh\tau},\\
        x_1&= \frac{R\rho \sinh\upsilon \cos\phi}{\rho\cosh\upsilon +\sqrt{\rho^2-1}\cosh\tau},\hspace{2cm}\xi = \frac{R\rho \sinh\upsilon \sin\phi}{\rho\cosh\upsilon +\sqrt{\rho^2-1}\cosh\tau},
    \end{split}
\end{align}
which after a Wick rotation brings us to the seven-dimensional hyperbolic black hole \cite{Casini:2011kv}
\begin{align}\label{eq:ads7-hypBH}
    ds_7^2 = \frac{d\rho^2}{f(\rho)} +f(\rho)d\tau^2 + \rho^2 d\upsilon^2 +\rho^2\sinh^2\upsilon (d\phi^2 + \sin^2\phi d\Omega_3^2)
\end{align}
where the metric function
\begin{align}
    f(\rho) = \rho^2 -1\,,
\end{align}
and $\tau \sim \tau + 2\pi$.  The BPS probe brane embedding found in \sn{embedding} now wraps $\{\tau,\upsilon,\rho\}$.  The probe brane contributions to the thermal entropy in the hyperbolic black hole background corresponds to the holographic computation of the BPS surface defect contribution to the EE of a spherical region in the dual field theory.

More generally, one can consider (not necessarily BPS) probe brane embeddings in seven-dimensional black hole solutions with metric function
\begin{align}
    f(\rho) = \rho^2 - 1 - \frac{\rho_h^4}{\rho^4}(\rho_h^2-1)
\end{align}
with the location of the horizon determined in terms of the inverse temperature $\beta$ by
\begin{align}\label{eq:rhoh}
    \rho_h = \frac{1}{3\beta}(\pi + \sqrt{\pi^2 +6\beta^2}).
\end{align}
This background is particularly useful for the holographic computation of R\'enyi entropies \cite{Hung:2011nu}.  Moving to a branched cover of the black hole geometry $\tau \sim \tau + \beta$ with $\beta = 2\pi n$, the thermal entropy in this background is dual to the $n^{\rm th}$-R\'enyi entropy
\begin{align}\label{eq:renyi-entropy}
    S^{(n)} = \frac{2\pi n}{1-n}(\CF(2\pi) - \CF(2\pi n)),
\end{align}
where $\CF(\beta) = \beta^{-1}S_{\rm ren}(\beta)$.  Clearly from the bulk geometry, the limit $n\to 1$ recovers the seven-dimensional hyperbolic black hole background, and in the dual field theory $S^{(n)}|_{n\to1}\to S_{\rm EE}$ \cite{Hung:2011nu,Jensen:2013lxa}, which holographically 
\begin{align}\label{eq:SEE-dbeta-Sren}
    S_{\rm EE} = \beta^2 \pd_\beta (\beta^{-1} S_{\rm ren}(\beta))|_{\beta\to 2\pi} \,.
\end{align}

Computing the holographic renormalized on-shell action, we begin with 
\begin{align}
    \vol{\tilde M_3} = \int_0^\beta d\tau\ \int_0^{\upsilon_c} d\upsilon\int_{\rho_h}^\Lambda\rho d\rho = \frac{\beta}{2}\upsilon_c(\Lambda^2-\rho_h^2).
\end{align}
where we have introduced a large radial cutoff $\Lambda$. The $\Lambda^2$ divergence is regulated by \eq{counterterm-action}
\begin{align}
    S_{\rm CT} = -\frac{32\pi}{81}\beta T_4\upsilon_c\Lambda^2\rho_h^2\alpha^j\,.
\end{align}
 Using \eq{CHM-map}, the linear divergence in $\upsilon_c$ in the black hole background is related to a log divergence at small $u=-\epsilon$
\begin{align}
    \upsilon_c = \log\frac{2R}{\epsilon}\,,
\end{align}
 where $R$ is the radius of the boundary spherical entangling surface that anchors the homologous Ryu--Takayanagi surface \cite{Ryu:2006bv} in the bulk.  Thus, we find that the near boundary behavior of the holographically renormalized action is\footnote{For $\beta\neq2\pi$, the defect embedding is expected to get deformed; this deformation is not known analytically and would not be BPS. However, we are evaluating the action, which on shell is stationary under small deformations; so $\partial_\beta S_{\rm ren}(\beta)|_{\beta\to 2\pi}=0$, and we can use directly the expression \eqref{eq:S-ren-EE} in \eqref{eq:SEE-dbeta-Sren} \cite{Rodgers:2018mvq,Kumar:2017vjv}.}
\begin{align}\label{eq:S-ren-EE}
    S_{\rm ren} = -\frac{8}{81}q_j\alpha^j T_4 \vol{\mathds{S}^2}\beta \rho_h^2 \log\frac{2R}{\epsilon} +\ldots\,.
\end{align}
 Using the form of $\rho_h$ in \eq{rhoh}, we find the defect sphere EE is
\begin{align}\label{eq:defect-EE}
    S_{\rm EE} = \frac{8}{405\pi^2}q_j\alpha^j \log \frac{2 R}{\eps}+\ldots\,.
\end{align}
In the $F_0=0$ case, we can evaluate the previous quantities more explicitly using \eq{massless}. For example \eq{defect-EE}
becomes 
\begin{align}\label{eq:defect-EE-massless}
    S_{\rm EE} = \frac{4{k}}{5}\sum_j q_j j(N-j)\log\frac{2R}{\epsilon}+\ldots.
\end{align}

From \cite{Kobayashi:2018lil, Jensen:2018rxu}, we know that for a flat two-dimensional conformal defect embedded in a CFT on a flat six-dimensional background the universal part of the defect sphere EE is uniquely determined by defect Weyl anomalies:
\begin{align}\label{eq:defect-EE-anomalies}
    S_{\rm EE} = \frac{1}{3}\left(a_\Sigma - \frac{3}{5}d_2\right)\log\frac{2R}{\epsilon}
\end{align}
Further, we know that for a two-dimensional superconformal preserving at least $\CN=(2,0)$ defect supersymmetry, the two B-type defect Weyl anomaly coefficients, $d_1$ and $d_2$, are equal \cite{Bianchi:2019sxz,Drukker:2020atp}, and so the universal part of the defect sphere EE, when combined with an independent computation of either $a_\Sigma$ or $d_2$, provides some measure of characterization of the defect degrees of freedom via anomalies. In the next subsection, we will compute the on-shell action for probe branes wrapping $\mathds{S}^2$ and $\mathds{S}^1\times\mathds{S}^1$ in the boundary geometry, which will give us independent predictions for $a_\Sigma$ and $d_2$. 

In addition to the defect sphere EE, which can be found by taking $n\to 1$ in \eq{renyi-entropy}, there are other limits of $S^{(n)}$ that contain interesting physical information (see e.g. \cite{Hung:2011nu}). By setting $\beta = 2\pi n$ and computing the renormalized on-shell action, we find that the $n^{\rm th}$ R\'enyi entropy is 
\begin{align}
    S^{(n)} =  T_4 q_j  \alpha^j\vol{\mathds{S}^2}\vol{\tilde{\mathds{S}}^2}\log\frac{2R}{\epsilon} \frac{2 (1-6n^2 +\sqrt{1 + 24n^2})}{729 n (1-n)}+\ldots.
\end{align}
Taking the $n\to 0$ limit counts the number of non-zero eigenvalues of the reduced density matrix:
\begin{align}
    \lim_{n\to 0} S^{(n)} = \frac{4}{729n\pi^2}q_j\alpha^j \log \frac{2R}{\epsilon} +O(n^0)\,.
\end{align}
Taking the opposite limit, $n\to\infty$, measures the largest eigenvalue of the reduced density matrix
\begin{align}
    \lim_{n\to\infty}S^{(n)} = \frac{4}{243\pi^2}q_j\alpha^j\log\frac{2R}{\epsilon} + O(n^{-1})\,.
\end{align}

\subsection{On-shell action}
In this subsection, we compute the holographically renormalized on-shell action in \eq{renormalized-os} starting from \eq{D4-probe-action-OS} for the cases where the boundary geometry of $\tilde M_3$ is $\mathbb{R}^2$, $\mathds{S}^2$, and $\mathds{S}^1\times\mathds{S}^1$.  In the latter two cases, the computations result in robust predictions for physical quantities that will be compared to anomaly inflow and supersymmetric localization at large $N$ in the following sections.

\subsubsection{\texorpdfstring{$\mathbb{R}^2$}{R2}}

Here we compute the on-shell action for a probe D4-brane embedding into \eq{ads7} wrapping AdS$_3\times\mathds{S}^2$ engineering a flat two-dimensional defect supported on $\mathbb{R}^{2}$ in the dual field theory defined on a $\mathbb{R}^{6}$ background. We take the AdS$_7$ to be written in flat slicing as in \eq{flat-sliced-ads7}.  We fix static gauge with $\xi=0$ such that the full ${\rm SO}(4)_N$ normal bundle symmetry is preserved. Computing the pullback onto the worldvolume of the probe D4-branes, the induced metric takes the form
\begin{align}\label{eq:worldvolume-metric}
    \frac{ds_\Sigma^2}{\pi\sqrt{2}} =f_1^2(dx_0^2 + dx_1^2) +f_2^2du^2 + f_3^2 d\Omega_2^2.
\end{align}

Computing the on-shell action \eq{D4-probe-action-OS}, only finding $\vol{\tilde M_3}$ remains to be done.  For the embedding described above,
\begin{align}
    \vol{\tilde M_3} = \vol{\mathbb{R}^{2}}\int_{-\infty}^{-\epsilon}\frac{du}{u^3}~ = \frac{\vol{\mathbb{R}^{2}}}{2\epsilon^2},
\end{align}
where we have introduced a small radial cutoff $u = -\epsilon$.  There are no subleading divergences, and so the covariant counterterm action built out of the induced volume of the cutoff slice $u=-\epsilon$ exactly cancels $S_{\rm probe}$ leaving
\begin{align}
    S_{\rm ren} = 0\,,
\end{align}
as expected.

\subsubsection{\texorpdfstring{$\mathds{S}^2$}{S2}}

Since we are studying a particularly simple D4--D2 bound state system as a probe, we can easily consider other AdS$_7$ and AdS$_3$ geometries that will reveal the anomaly.  Instead of using flat slicing in Poincar\'e coordinates, we can consider the background AdS$_7$ geometry in global coordinates with radial slices having $\mathds{H}^5 \times\mathds{S}^1$ geometry using the map in \app{conformal}, which results in
\begin{align}\label{eq:AdS7-metric-H5-S1}
    ds_7^2 = d\sigma^2 + \cosh^2\sigma ds_{\mathds{H}_5}^2 +\sinh^2 \sigma d\varphi^2\,,
\end{align}
where the line element on the $\mathds{H}_5$ is given in \eq{AdS7-metric-H5-S1-unpacked}. So \eqref{eq:AdS7-metric-H5-S1} is essentially \eqref{eq:ads7-hypBH}, but with the defect now wrapping an equatorial $\mathds{S}^2\hookrightarrow\mathds{S}^4\subset\mathds{H}_5$. Thus, our D4--D2 bound state probe can be embedded in such a way that it wraps the boundary submanifold $\psi = \pi/2$ at the boundary $\rho\to\infty $ at fixed $\sigma$.

Computing the on-shell action in this background is straightforward:
\begin{align}
    S_{\rm probe} = \frac{16}{81}T_4(4\pi)^2\alpha^j\int_{0}^{\log\Lambda}d\rho \sinh^2\rho = \frac{32\pi^2}{81}T_4\alpha^j\left(\Lambda^2 -4\log\Lambda +\ldots\right)\,.
\end{align}
We have introduced a radial cutoff $\rho=\log \Lambda \gg 1$ and suppressed further subleading (finite) terms.  Note that in this curved space embedding, $z=j$ still extremizes the action, and so we can use this to evaluate the action on-shell.  The counterterm action is computed from \eqref{eq:counterterm-action}.  We can tune $c_1$ to remove the $\Lambda^2$ divergence, and so the holographically renormalized on-shell action is
\begin{align} \label{Sren}
  S_{\rm ren} = -\frac{8}{81\pi^2}q_j\alpha^j \log \Lambda\,.
\end{align}
This log divergence is physical: it holographically computes the A-type Weyl anomaly for a defect wrapping $\mathds{S}^2$  inserted into the six-dimensional $\CN=(1,0)$ SCFT dual to the ten-dimensional SUGRA solution above:
\begin{align}
    \left< \CO_{\mathds{S}^2}\right> \approx e^{-S_{\text{ren}}} = (R\mu)^{a_\Sigma/3} 
\end{align}
where we have introduced the length scale $R$ on the $\mathds{S}^2$ along with the RG scale $\mu$ for dimensional reasons. Equating 
\bes{
S_{\text{ren}} = - \frac{1}{3} a_\Sigma \log \Lambda
}
and using \eqref{Sren}, we find a prediction for the A-type defect Weyl anomaly to leading order in large $N$:
\begin{align}\label{eq:a-sigma}
    a_\Sigma = \frac{8}{27\pi^2}q_j\alpha^j\,.
\end{align}
In order to compute the B-type anomaly, we insert \eqref{eq:a-sigma} into \eqref{eq:defect-EE-anomalies} and use \eqref{eq:defect-EE} to arrive at
\begin{align}
   d_1= d_2 = \frac{32}{81\pi^2}q_j\alpha^j\,.
\end{align}

We can express $a_\Sigma$ in terms of representation theoretic data, like for the Wilson surface in the $(2,0)$ theory \cite{Estes:2018tnu, Jensen:2018rxu}. Recalling \eq{a-Cr}, we can rewrite
\begin{equation} \label{eq:gravres}
        a_\Sigma = 24 q_i (C^{-1})^{ij} r_j = 24 (q,r) 
        \,,\qquad 
        d_1=d_2 = 32(q,r)\,,
\end{equation}
where we defined the Killing form on the weight space:
\begin{equation}
    (\lambda,\lambda^\prime) \equiv \lambda_i (C^{-1})^{ij}\lambda^\prime_j\,.
\end{equation}

As a simple example, consider the case where $F_0=0$. All the $r_i=k$, so $r=k \rho_W$, where $\rho_W=(1,1,\ldots,1)$ is the the Dynkin label for the $A_{N-1}$ Weyl vector. Moreover take a single probe brane in position $j$: $q_j=1$, $q_{i\neq j}=0$. We obtain
\begin{align}
    a_\Sigma = 12{k}\sum_j j(N-j) = 24{k} (\lambda,\,\rho_W) = \frac{3}{4}d_2\,
\end{align}
where now
\begin{align}
    \lambda = (\underbrace{0,\ldots,0}_{j-1},1,\underbrace{0,\ldots,0}_{N-j-1})\,.
\end{align}

\subsubsection{\texorpdfstring{$\mathds{S}^1\times\mathds{S}^1$}{S1xS1}}\label{sec:probe-s1xs1}

Even though we already have the two independent defect Weyl anomalies in hand, there are other defect geometries whose partition function carry non-trivial information related to anomalies. Here we consider the probe brane geometry dual to the Wilson surface operator wrapping $\mathds{S}^1_{R_6}\times\mathds{S}^1\hookrightarrow\mathds{S}^1_{R_6}\times\mathds{S}^5$.
The purpose of computing the on-shell action for this probe brane configuration is two-fold.  First, upon dimensional reduction along $\mathds{S}^1_{R_6}$, the BPS Wilson surface operator descends to the circular Wilson loop $W$ in $\mathcal{N}=1$ SYM theory on $\mathds{S}^5$.  This opens up the possibility of comparing to exact results for $\left<W\right>$ obtained in the field theory using supersymmetric localization.  Second, the partition function $Z$ of an SCFT on $\mathds{S}^1_\beta\times\mathds{S}^{d-1}$ has long been known to be related to the superconformal index $\mathcal{I}$  up to an exponential pre-factor containing a term known as the supersymmetric Casimir Energy (SCE), $Z = e^{-\beta E_c}\CI~$ that can be computed in terms of various anomalies \cite{Assel:2014paa,Assel:2015nca,Bobev:2015kza}. In the case of a superconformal field theory, the defect SCE can be related to Weyl anomalies and in the case of superconformal defects, has been computed in several cases \cite{Chalabi:2020iie}. However, it only has been rigorously proven to be related to defect anomalies for co-dimension four defects in six dimensional $\CN=(2,0)$ SCFTs \cite{Meneghelli:2022gps}.

Returning the probe brane computation, the coordinate transformation to take us to a probe brane wrapping $\mathds{S}^1\times\mathds{S}^1$ is discussed in \app{conformal} ending with the AdS$_7$ metric in \eq{ads7-s1-s5-metric}.  We then take the periodicity around the $\mathds{S}^1_{R_6}$ to be $\tau \sim \tau + 2\pi R_6$.  On the field theory side in the dimensional reduction along $\mathds{S}^1_{R_6}$ to a Wilson loop wrapping a great circle on $\mathds{S}^5$, the five-dimensional Yang-Mills coupling is related to the radius of the compactified direction by
\begin{align}
    R_6 = \frac{g_{\rm YM}^2}{8\pi^2}\,.
\end{align}
Integrating over the $\mathds{S}^2$ and the $\mathds{S}^1$s wrapped in the boundary geometry, the on-shell action takes the form
\begin{align}
    S_{\rm probe} = \frac{128\pi^3R_6}{81}T_4q_j\alpha^j\int_0^{\log\Lambda}d\sigma \sinh 2\sigma  = -\frac{64\pi^3R_6}{81}T_4q_j\alpha^j\left(\frac{\Lambda^2}{2} -1 +\frac{1}{2\Lambda^2}+\ldots\right)
\end{align}
Again using \eq{counterterm-action} as the counterterm action, the $\Lambda^2$ divergence is regulated, and we find the holographically renormalized on-shell action
\begin{align}\label{eq:s1-s1-defect-ren-action}
   S_{\rm ren} = -\frac{64\pi^3R_6}{81}T_4  q_j\alpha^j\,.
\end{align}
In order to more easily compare to the field theory computation, we define $\beta = 4\pi R_6$, hence
\begin{align}
    \left< W\right> = \exp\left[\frac{\beta}{81\pi^2}q_j\alpha^j\right] = \exp\left[\beta q_i (C^{-1})^{ij}r_j\right]\,,
\end{align}
which in the $F_0=0$ case gives
\begin{align}\label{eq:holographic-Wilson-surface-vev}
    \left<W\right> =  \exp\left[\frac{\beta}{2}{k}\sum_j q_jj(N-j)\right].
\end{align}
In the following section, we will compare \eq{holographic-Wilson-surface-vev} to the field theory calculation of the expectation value of a circular Wilson loop in five-dimensional SYM theory on $\mathds{S}^5$ computed using supersymmetric localization.

Due to the nature of the holographic computation, it should not be entirely surprising that $\left<W\right>$ can be expressed in a form similar to the defect Weyl anomalies. However, without detailed knowledge of how the defect SCE of surface operators in six-dimensional $\CN=(1,0)$ SCFTs should be related to defect Weyl anomalies, we cannot say what linear combination of $a_\Sigma$ and $d_2$ should be computed by $\left<W\right>$.  In the case of a Wilson surface in the six-dimensional $A_{N-1}$ $\mathcal{N}=(2,0)$ theory, however, it has been been shown using chiral algebra methods that $d_2$ alone appears in the defect SCE \cite{Meneghelli:2022gps}, which leads to the conjecture  that the relation to the defect Weyl anomalies in \eq{holographic-Wilson-surface-vev} is not just superficial and that at large $N$ the defect SCE is given by $E_c =  d_2/32$, though a proof is lacking.

\subsection{Defect gravitational anomaly} 

Here, we compute the defect gravitational anomaly from the worldvolume action of the probe brane in the AdS$_7 \times M_3$ geometry in \eq{ads7}. In the anomaly polynomial this is the coefficient of the first Pontryagin class of the defect submanifold $\Sigma$:
\begin{equation}\label{eq:a-kg}
    \mathcal I_4 =\frac{k_r}{2}c_1(r) - \frac{k_g}{24}p_1(T)\,.
\end{equation}
$p_1(T\Sigma)=\frac{{\rm Tr}(R_T \wedge R_T)}{8\pi^2}$ is related by descent mechanism to the gravitational three-dimensional Chern--Simons form,
\begin{equation}
    p_1 = 2 \pi d  {\rm CS}_3(R_T)\,\qquad
    {\rm CS}_3(R_T) = \frac{1}{4 \pi} {\rm tr} \left(\Gamma_T \wedge d \Gamma_T + \frac{2}{3}\Gamma_T^3 \right) \, ,
\end{equation}
where ${R_T}^a{}_b=(d\Gamma_T+ \Gamma_T^2)^a{}_b =\frac{1}{2}R^{a}{}_{b\mu \nu} dx^{\mu}\wedge dx^{\nu}$ is the curvature of $\Sigma$. Explicitly we have 
\begin{equation} \label{eq:Rtrace}
{\rm tr} (R_T\wedge R_T)= \frac{1}{4} R^{a}{}_{b\rho \sigma} R^{b}{}_{a\mu \nu} dx^{\mu}\wedge dx^{\nu}\wedge dx^{\rho}\wedge dx^{\sigma}
\end{equation}
The anomaly appears in the partition function as a phase,
\begin{equation} \label{eq:anact}
    e^{2 \pi i \mathcal{A}} = e^ {i k_g \int_{M_3} \frac{\text{CS}_3(R_T)}{24}}
\end{equation}
where for us $M_3$ will be AdS$_3$.

In order to evaluate the gravitational anomaly, we apply the strategy adopted in \cite{Aharony:2007dj}. To start, we need to look at the worldvolume action of a stack of $q_i$ D4-branes wrapping $\mathds{S}^2_{z=i}\subset M_3$ and extending along AdS$_3$. In particular we focus on the Wess--Zumino term that includes the gravitational contribution,
\begin{equation} \label{eq:CSgrav}
    S_{\rm D4}^{\rm WZ}= T_4 \int_{M_8} {\rm Tr} \left[ \sum_n C_n \wedge e^{\cal F} \wedge \sqrt{\hat{A}(2\pi R_T)}\right]_5\,,
\end{equation}
where the A-roof genus reads
\begin{equation}
    \hat{A}(R) = 1 - \frac1{24}p_1(T\Sigma) + \ldots\,.
\end{equation}
If we ignore the term due to the non-abelian scalars, the term proportional to $p_1(T\Sigma)$ reads
\begin{equation}
    S_{\rm D4}^{\rm WZ} \supset-  q_i\int_{M_5}  (C_1 - C_{-1} B_2) \wedge \frac{1}{24}p_1(T\Sigma)\,,
\end{equation}
where $q_i$ comes from the trace of the identity matrix. In addition we also accounted for the IIA Romans mass contribution $F_0$ which formally can be written as $F_0=dC_{-1}$. Integrating by parts and by using \eqref{eq:CSgrav}, we obtain
\begin{equation}
    S_{\rm D4}^{\rm WZ} \supset -q_i\int_{\mathds{S}^2_{z=i}}  (F_2 - F_0 B_2)  \int_{\text{AdS}_3} \frac{{\rm CS}_3(R_T)}{24
\times 2 \pi}\, .\end{equation}
By inputting the \eq{ads7} and by integrating over $\mathds{S}^2_{z=i}$ we get
\begin{equation}\label{eq:defect-gravitational-anomaly}
      S_{\rm D4}^{\rm WZ} \supset-  q_i r_i \int_{\text{AdS}_3} \frac{{\rm CS}_3(R_T)}{24}\,.
\end{equation}
We now have to compare this result with the argument in \eqref{eq:anact}.
Considering multiple stacks of D4-branes, we obtain the gravitational anomaly
\begin{equation}
    k_g = -  \sum_i q_i r_i\,.
\end{equation}
This matches the field theory computation in \eq{kg}.

Let us now briefly comment on the computation that led to \eq{defect-gravitational-anomaly} with an eye toward \sn{inflow}. First, notice that the defect gravitational anomaly is not directly captured by inflow. Further, it does not have the same product structure of the bulk-defect degrees of freedom, where the metric of the string lattice and its dual appear. A possible interpretation for this is that the gravitational anomaly accounts for the modes living only on the defect and not interacting with the bulk six-dimensional SCFT. The latter does not capture the inflow contribution, but it reproduces the gravitational anomaly computation that we just performed in gravity purely from field theory. Finally we speculate that the gravitational anomaly computation provides  the only contribution to the subleading correction of the defect $a_\Sigma$ anomaly in the large $r_i$ and $N$ limit. In other words, the on-shell action provides the leading contribution, whereas the WZ action gives the subleading one. This would work exactly like in \cite{Aharony:2007dj}. The total $a_\Sigma$ anomaly would then read \cite{Wang:2020xkc}
\begin{equation}\label{eq:a-sigma-grav}
    a_\Sigma = 3 k_r - \frac{k_g}{2} = 24 q_i (C^{-1})^{ij}r_j+ \frac{\sum_i q_i r_i}{2}\,.
\end{equation}

\section{Anomalies from Field Theory and their string theory constructions}\label{sec:ft-anomalies}

In this section, we test the holographic predictions provided by the probe brane computations in the previous section in two different settings.  In the first subsection, we employ supersymmetric localization to compute the expectation value of a circular Wilson loop operator inserted in the antisymmetric representation of a necklace quiver gauge theory on $\mathds{S}^5$. This  Wilson loop operator is thought to have UV completion to the string defect operator studied in \sn{probe-s1xs1}, and so we compare the localization computation to the holographic result for $\left<W\right>$. In the second subsection, we take a different approach. We first rely on IIB/F-theory constructions of the 6d $\mathcal{N}=(1,0)$ SCFTs to compute defect anomalies using inflow arguments. We then use known non-perturbative relations between 't Hooft anomalies and the A-type Weyl anomaly for superconformal two-dimensional defects to compute $a_\Sigma$.  In both cases, we find exact agreement with the probe brane results.

\subsection{Necklace quivers and supersymmetry localization on \texorpdfstring{$\mathds{S}^5$}{S5}}\label{sec:loc}
In this section, we demonstrate that the holographically renormalized on-shell action in \eq{s1-s1-defect-ren-action} can be matched with the expectation value of the Wilson loop in the five-dimensional $\CN=1$ theory described by a circular quiver with $k$ $\mathrm{SU}(N)$ gauge groups and the massless bifundamental hypermultiplet between the two gauge groups that are next to each other. Here the Wilson loop is taken to be in the rank-$j$ antisymmetric representation of each gauge group. Such a quiver theory can be regarded as a Kaluza--Klein (KK) theory whose UV completion is the six-dimensional $\CN=(1,0)$ theory living on $N$ M5-branes on the $\mathbb{C}^2/\mathbb{Z}_k$ singularity (see \eg~ \cite[Sec. 4]{DelZotto:2014hpa}). Just like in Sec.~\ref{sec:ads7}, this six-dimensional $\CN=(1,0)$ theory can be realized using the Type IIA brane system consisting of $k$ D6-branes in the $0123456$ directions, along with $N$ NS5-branes in the $012345$ directions; unlike in the general case, no D8-branes are necessary. The defect is regarded as a D4-brane spanning the $01789$ directions and is located between the $(j-1)$-th and $j$-th NS5-branes. In particular, we will show that the expectation value of the Wilson loop in question is related to the Weyl anomalies of the D4-brane defects in such a brane configuration. 

The expectation value of the Wilson loop in the rank-$j$ antisymmetric representation $\wedge^j$ of five-dimensional $\CN=1$ SYM with $\mathrm{SU}(N)$ gauge group was performed in \cite[Sec. 2.3]{Mori:2014tca}:
\begin{align} \label{eq:expWilsonAk}
\left< W_{\wedge^j} \right> = \exp \left(\frac{\beta}{2} j(N-j) \right)\,, \qquad \beta \equiv \frac{g_{\text{YM}}^2}{2 \pi r}
\end{align}
where $r$ is 
the radius of the five-sphere which the Wilson loop wraps.
On the other hand, the authors of \cite{Minahan:2013jwa, Minahan:2016xwk} also considered the circular quiver as described above, where the Wilson loop was taken to be in the fundamental representation $\mathbf{N}$ of {\it one} of the $\mathrm{SU}(N)$ gauge groups and in the trivial representation $\mathbf{1}$ of the others. They found that its expectation value turns out to be equal to \eq{expWilsonAk} with $j=1$, namely that of the Wilson loop in the fundamental representation of the five-dimensional $\CN=1$ SYM with $\SU(N)$ gauge group \cite{Kim:2012qf, Minahan:2013jwa}:
\begin{align}
\left< W_{(\mathbf{N},\mathbf{1},\ldots, \mathbf{1})} \right> = \exp \left(\frac{\beta}{2} (N-1) \right) \overset{N \rightarrow \infty}{\sim}  \exp \left(\frac{\beta}{2} N \right)\,.
\end{align}
where the gauge coupling of each gauge group in the circular is taken to be equal to $g_{\text{YM}}$. Our main goal is to generalize these results, namely to compute the expectation value $\langle W_{(\wedge^j,\wedge^j,\ldots, \wedge^j)} \rangle$ of the Wilson loop in the rank-$j$ antisymmetric representation of every $\mathrm{SU}(N)$ gauge group in the circular quiver. 

To achieve this, we proceed as in \cite[Sec. 4.3]{Minahan:2013jwa}, \cite[Sec. 3]{Minahan:2016xwk} and \cite[Sec. 2]{Mori:2014tca}. According to \cite[(3.22)]{Minahan:2016xwk}, the eigenvalue distribution of each $\mathrm{SU}(N)$ gauge group in the circular quiver can be taken to be the same. We denote these by $\phi_i$, with $i=1, \ldots, N$. The expression of the partition function then simplifies, whereby in the large $N$ limit it takes the form
\begin{align} \label{eq:Zcirc}
Z \sim \int \prod_{i=1}^N d\phi_i \, \exp\left[ - k \frac{N^2}{\beta} \sum_{i=1}^N \phi_i^2 +k \frac{N}{2} \sum_{1\leq i \neq l \leq N} |\phi_i-\phi_l| \right]
\end{align}
which is indeed a simple modification of the matrix model associated with five-dimensional $\CN=2$ SYM given by \cite[(2.3)]{Mori:2014tca}. The latter can simply be obtained from \eq{Zcirc} by setting $k=1$. The saddle points are the same as in \cite[(2.4), (2.5)]{Mori:2014tca}:
\begin{align}
0 &= -\frac{2N^2}{\beta} \phi_i + N \sum_{l, i \neq l} \mathrm{sign}(\phi_i -\phi_l) \,;\\
\phi_i &= \frac{\beta}{2N} (N-2i)\,, \quad \text{for $\phi_i > \phi_l$ whenever $i<l$.} \label{saddles}
\end{align}
Note that, upon evaluating the exponential function in the integrand of \eq{Zcirc} at the saddle points, we find that the free energy $F=-\log Z$ obtained from above is indeed $k$ times that of the five-dimensional $\CN=2$ SYM with $\mathrm{SU}(N)$ gauge group, as pointed out in \cite[(3.11)]{Kallen:2012zn}, \cite[(4.30)]{Minahan:2013jwa}, and \cite[Sec. 4.3]{Minahan:2016xwk}.  Let us now proceed with the computation of the Wilson loop expectation value. Similarly to \cite[(2.15)]{Mori:2014tca}, this is given by
\begin{align}
\left< W_{(\wedge^j,\wedge^j,\ldots, \wedge^j)} \right> \sim \int \prod_{i=1}^N d\phi_i \, \exp\left[ - k \frac{N^2}{\beta} \sum_{i=1}^N \phi_i^2 +k \frac{N}{2} \sum_{1\leq i \neq l \leq N} |\phi_i-\phi_l| + k N \sum_{i=1}^j \phi_i \right]\,.
\end{align}
Since the derivative with respect to $\phi_i$ of the last term in the exponential gives a constant $kN$ for all $i=1, \ldots, j$, the insertion of the Wilson loop does not change the eigenvalue distribution. Evaluating the exponential function containing the last term yields
\begin{align}
\left<W_{(\wedge^j,\wedge^j,\ldots, \wedge^j)} \right> \sim
\exp \left(k N \sum_{i=1}^j \phi_i \right) \Bigg |_{(\ref{saddles})} \sim \exp \left(\frac{\beta}{2} k j(N-j) \right)\,.
\end{align}
The argument of the exponential function is in agreement with the holographic prediction for the surface operator wrapping $\mathds{S}^1\times\mathds{S}^1$ with vanishing Romans mass in \eq{holographic-Wilson-surface-vev} for a single defect $q_j =1$.

\subsection{Anomaly Inflow from type IIB}\label{sec:inflow}
In this subsection, we derive the $a_\Sigma$ anomaly on the defect via inflow mechanism from the F-theory construction of six-dimensional $\mathcal N={(1,0)}$ SCFTs \cite{Heckman:2013pva, DelZotto:2014hpa, Heckman:2015bfa}. These backgrounds can be described in terms of type IIB string theory on a complex two-dimensional space with a non-trivial fibration of the axio-dilaton. When the space is singular, IIB on the singularity engineers the SCFT. If instead the space is resolved, which then denote by $\mathcal B$, IIB provides a definition of the tensor branch effective field theory (EFT) and corresponds to the scalars in the tensor multiplets acquiring non-trivial vacuum expectation values. The full BPS physics of the tensor branch EFT is determined by the geometric structure of the 2-cycles in $\mathcal B$ which read
\begin{equation} \label{eq:intpair}
\begin{aligned}
    & \int_{\mathcal B} \beta_i \wedge \beta_j = A_i \cdot A_j =\int_{A_i}\beta_j =  \Omega_{ij}\\
    & \int_{\mathcal B} \beta_i \wedge \alpha^j = A_i \cdot B^j =\int_{B^j}\beta_i= \delta_{i}^{j}\,.
\end{aligned}
\end{equation}
Here $\cdot$ denotes the number of intersections between cycles, weighted by signs so as to be topologically invariant; $\beta_i,\alpha^i,B^i,A_i$ generate the compact and non-compact cohomology and homology:
\begin{equation}
\begin{aligned}
 &   \beta_i \in H^2_{\rm c}(\mathcal B), \quad \alpha^i \in H^2_{\rm nc}(\mathcal B)\\
 &   B^i \in H_2^{\rm nc}(\mathcal B), \quad A_i \in H_2^{\rm c}(\mathcal B)\,.
\end{aligned}
\end{equation}
The axio-dilaton depends on the coordinates of the base and it is interpreted as the complex structure of a $\mathds{T}^2$ fibered over $\mathcal{B}$. Crucially, the axio-dilaton degenerates over some loci of 
$\mathcal{B}$, and the type of monodromy around these singular loci determines which D7-branes spans the singular loci. Since we want to have supersymmetric configurations, the singular loci themselves are the holomorphic 2-cycles of $\mathcal{B}$. In addition, we associate a gauge algebra to a given D7-brane, which can be trivial or non-trivial or even of exceptional type. The RR four-form potential of type IIB is expanded as 
\begin{equation}
    C_4=B_2^i \wedge \beta_i 
\end{equation}
providing the dynamical anti-symmetric two-form fields $B_2^i$. The bosonic content of the tensor multiplet is then completed by the real scalar modulus corresponding to
\begin{equation}
    {\rm Vol} (A_i) = \int_{A_i}J = \Omega_{ij} \phi^j, \qquad J= \phi^i \beta_i\,,
\end{equation}
where $J$ is the K\"ahler form of $\mathcal B$.
Lastly, we have D3-branes wrapping $A_i$, which are electrically charged under $C_4$ and therefore under the tensor multiplets. They form a lattice of BPS strings whose pairing is given by $\Omega_{ij}$. 

This lattice of string operators plays a crucial role in determining the spectrum of defect operators. In a four-dimensional gauge theory the ``defect group'' $\Lambda^*_{\rm root} / \Lambda_{\rm root}$ (where $\Lambda_{\rm root}$ is the root lattice of the gauge algebra and $\Lambda^*_{\rm root}$ is its dual) organizes the set of charged operators that cannot be screened.  On a basic level, the dual lattice tells us about the non-trivial line defects that the theory supports. The geometric understanding of the origin of these lattices from the brane picture (at least for SCFTs) can be lifted to a similar defect group for six-dimensional theories \cite{DelZotto:2015isa}, where now  $\Lambda_{\rm string}$ is the charge lattice of tensionless BPS strings and $\Lambda^*_{\rm string}$ is the lattice of string defects that the theory supports. The six-dimensional defect group is then given by $\Lambda^*_{\rm string}/\Lambda_{\rm string}$, which determines the strings charged under self-dual two-forms that cannot be screened.  

\subsubsection{Anomaly Polynomial}
The anomaly polynomial, $I_8$, capturing the perturbative 't Hooft anomalies for continuous symmetries of six-dimensional $\mathcal N={(1,0)}$ theories has been well studied in \cite{Cordova:2015fha,Ohmori:2014kda,Intriligator:2014eaa} and can be computed from the low-energy EFT on the tensor branch.  The necessary condition is that the tensor branch EFT contains the appropriate Green--Schwarz (GS) coupling to cancel reducible gauge anomalies. That is, $I_8 \sim (I_4^{\rm gauge})^2$ where $I_4^{\rm gauge}= 1/2 {\rm tr}_{\lambda} (F_g \wedge F_g)$, $F_g$ is the field strength of the gauge field, and $\lambda$ is a representation of the gauge algebra. Let us define the Dirac pairing of BPS strings,
\begin{equation}
    C_{ij}=-\Omega_{ij}\,,
\end{equation}
where $C_{ij}$ is the same intersection paring introduced in \eqref{eq:intpair}. Then the GS coupling \cite{Green:1984bx, Sagnotti:1992qw} takes the form 
\begin{equation} \label{eq:GScoupl}
    S_{\rm GS}= \int_{M_6} C_{ij} B_2^i \wedge I_4^j \,.
\end{equation}
$S_{\rm GS}$ couples the tensors to the gauge fields, but $I_4^i$ also contains characteristic classes of the global symmetries like the first Pontryagin class of the tangent bundle $p_1(T)$, as well as the the second Chern class of the ${\rm SU}(2)_{R_I}$-symmetry bundle $c_2(I)$. 

Let us be more specific:
\begin{equation} \label{eq:I4}
    I_4^i = \frac{1}{4} {\rm Tr}(F_i\wedge F_i) + y^i c_2(I) - K^i p_1(T)\,,
\end{equation}
where ${\rm Tr} \cong {\rm tr}_{\lambda}/{\rm Ind}(\lambda)$ is the normalized trace over the index of the representation $\lambda$, and $F_i$ is the field strength of the gauge field associated to the D7-brane wrapping $A_i$; keeping in mind that this can be the trivial gauge algebra as well. In addition we have that for a one-instanton background $\int_{\Sigma_4} \frac{1}{4} {\rm Tr}(F\wedge F) =1 $ for a closed $\Sigma_4 \subset M_6$. $y^i, K^i$ are coefficients such that also mixed gauge--global anomalies are canceled, this is due to the fact that six-dimensional SCFTs do not have two-form conserved currents \cite{Cordova:2020tij}. This fixes 
\begin{equation} \label{eq:yi}
    y_i = (C^{-1})^{ij}r_j, \qquad  r_i= h_{\mathfrak{g}_i}^{\vee}\,,
\end{equation}
where $h_{\mathfrak{g}_i}^{\vee}$ is the dual Coxeter number of the gauge algebra $\mathfrak{g}_i$. In addition,
\begin{equation}
    K^i = (C^{-1})^{ij} (2 - C_{jj})\,.
\end{equation}
The GS contribution to the anomaly polynomial then reads
\begin{equation} \label{IGSI4I4}
    I_{\rm GS}= \frac{1}{2}C_{ij}I_4^i I_4^j\,.
\end{equation}

The origin of $I_{\rm GS}$ from the IIB background was proposed in \cite{Ohmori:2014kda}. This comes in term of a formal twelve-form which we then integrate over the IIB base $\mathcal{B}$:
\begin{equation} \label{eq:IIBGS}
    I_{\rm GS}= - \frac{1}{2}\int_{\mathcal B} Z^2\,, \qquad dF_5=Z\,,
\end{equation}
where $F_5$ is the five-form flux, and \cite{Sadov:1996zm}
\begin{equation}
    Z= \frac{1}{4}\left(c_1(\mathcal{B}) \wedge p_1(T) + \sum_i {\rm Tr}(F_i\wedge F_i)\wedge \beta_i\right)\,.
\end{equation}
$c_1(\mathcal{B})$ is the first Chern class of the base, and we have the relation with the anticanonical bundle of $\mathcal B$, 
\begin{equation}
    \int_{\mathcal{B}} c_1(\mathcal{B}) \wedge \beta_i = C_{ij} K^j= (2-C_{ii}),
\end{equation}
along with
\begin{equation} \label{eq:flured}
    F_5 = H^i \wedge \beta_i\,, \qquad dH^i = I_4^i\,, \qquad -C_{ij} I^j = \int_{\mathcal{B}} Z \wedge \beta_i, \qquad Z=  I_4^i \wedge \beta_i  \,.
\end{equation}
We have assumed that $dH^i$ also contains a term proportional to $c_2(R)$, which is necessary for the mixed anomaly cancellation. On the other hand it is not directly visible from the IIB geometric construction because the R-symmetry is not manifest. Plugging \eqref{eq:flured} in \eqref{eq:IIBGS} we get \eqref{eq:GScoupl}.

The GS contribution can be also derived from the world-volume action of the D7-branes where the following term should be present: 
\begin{equation}
    \int_{M_6 \times A_i } C_4 \wedge I^i = \int_{M_6\times A_i } B_2^j \wedge \beta_j \wedge I^i =  \int_{M_6 } C_{ij} B_2^i  \wedge I^j 
\end{equation}
In case of D7-branes the contribution proportional to ${\rm Tr}(F_i\wedge F_i)$ and $p_1(T)$ come from the Wess--Zumino term,
\begin{equation}
    S_{D7}^{\rm WZ}= \mu_7 \int_{M_8} {\rm Tr} \left[ \sum_n C_n \wedge e^{\cal F} \right]_8\,,
\end{equation}
where we ignore the dependence on the scalar fields parameterizing the orthogonal directions.

\subsubsection{Anomaly inflow for BPS strings and surface defects}
As we anticipated, the D3-branes wrapping $A_i$ will provide the BPS strings charged under $B_2^i$. On the other hand, the D3-branes wrapping non-compact cycles $B^i$ realize the $\mathcal N=(0,4)$ surface defects in the theory \cite{DelZotto:2015isa}. Since both objects are D3-branes, they source additional terms in $F_5$:
\begin{equation} \label{eq:F5sourced}
        dF_5 = Z + Q^i \, (\delta^{(4)}(\mathbf{x}^{\perp}) d\mathbf{x}^{\perp} + \chi_4(N\Sigma)) \wedge \beta_i +  q_i\, (\delta^{(4)}(\mathbf{x}^{\perp}) d\mathbf{x}^{\perp} + \chi_4(N\Sigma)) \wedge \alpha^i\equiv Z'
\end{equation}
where $Q^i$ are the BPS string charges and $q_i$ are the number of defects. 
$\mathbf{x}^{\perp}$ are the coordinates of the space $\mathbb R^{\perp}_4 \subset M_6$ perpendicular to the BPS strings and surface defects in the six-dimensional space-time. 
The tangent bundle $TM_6$ of the background splits into the tangent bundle $T\Sigma$ and the normal bundle $N\Sigma$ (with structure group ${\rm SO}(4)$) of the defect submanifold $\Sigma$. So, as in \cite{Shimizu:2016lbw},  
\begin{equation}
\begin{aligned}
    &\chi_4(N\Sigma)=c_2(L)-c_2(R)\,, \\
    &p_1(TM_6) = p_1(T\Sigma)+ p_1(N\Sigma) = p_1(T\Sigma) - 2 (c_2(L)+c_2(R))\,.
\end{aligned}
\end{equation}

The surface defect is a D3 probe in the geometric F-theory setup, which in a string theory background can receive anomaly contributions from the bulk theory by inflow. Placing the probe brane in the string theory background produces a bulk anomaly for a given symmetry. This anomaly is then canceled by an anomaly for the probe brane \cite{Freed:1998tg,Kim:2012wc}. We can apply the same logic here. We need to integrate the anomaly polynomial in the presence of the D3 brane probe source given by \eqref{eq:F5sourced}. We can either start from IIB or directly from six dimensions. Only the reducible part will contribute $I_{\rm GS}$ to the inflow:
\begin{equation} \label{eq:IGSinfred}
    \int_{\mathbb R_4^{\perp}\times\mathcal{B}} I_{\rm GS}= -\int_{\mathbb R_4^{\perp}\times\mathcal{B}} \frac{1}{2} (Z')^2 \supset - \int_{\mathcal{B}} I_4^i \wedge \beta_i \wedge (Q^j  \beta_j + q_j \alpha^j) = C_{ij}I_4^i Q^j - q_j \delta_i^j I^i_4\,.
\end{equation}
The first term of \eqref{eq:IGSinfred} would now reproduce the one for the BPS strings computed in \cite{Shimizu:2016lbw}. But let us focus on the inflow on the defect only, neglecting the inflow contributions on the strings, setting $Q^i=0$. Plugging in \eqref{eq:I4} and \eqref{eq:yi}, 
\begin{equation}
    \mathcal I_4(Q^i=0)= - q_i (C^{-1})^{ij}r_j c_2(I)\,.
\end{equation}
We need to convert the ${\rm SU}(2)_{R_I}$ into the $(0,2)$ ${\rm U}(1)_r$ superconformal R-symmetry, and for this we use the result of \cite[(4.28)]{Wang:2020xkc},\footnote{Note that the second equality in \eqref{eq:yifanformula} is slightly different from \cite[(4.28)]{Wang:2020xkc}, since we are using half-integral charges with respect to $R,I$. We thus remove a factor of 2 in the formula \cite[(4.28)]{Wang:2020xkc}.}
\begin{equation} \label{eq:yifanformula}
   a_\Sigma =3k_r - \frac{k_g}{2}, \qquad r=(R-2I)\,,
\end{equation}
recalling the definition of $k_r$ in \eqref{eq:a-kg}.

The ${\rm SU}(2)_{R_I}$ vector bundle, $V_I$, splits in terms of its Cartan line bundle $L_I$ as $V_I=L_I\oplus L_I^{\vee}$. This implies that 
\begin{equation} \label{c2Iandc1r}
    c_2(I)= - c_1(L_I)^2 = -4 c_1(r)^2\,.
\end{equation}
The factor of 4 comes from \eqref{eq:yifanformula}; we ignored the contribution from $c_2(R)$ since it vanishes when $Q^i=0$. In the anomaly polynomial we have,
\begin{equation}
    K_I c_2(I) = - 4 K_I c_1(r)^2 = \frac{k_r}{2}c_1(r)^2, \qquad K_I= - q_i (C^{-1})^{ij}r_j\,,
\end{equation}
which implies that 
\bes{ \label{krKI}
k_r = - 8 K_I~
}
The inflow contribution to the surface defect $a_\Sigma$ anomaly of a generic six-dimensional SCFT is then
\begin{equation}\label{eq:aS-inflow}
    a_\Sigma=  24 q_i (C^{-1})^{ij}r_j\,.
\end{equation}
In the case where $C_{ij}$ is the Cartan matrix of $\mathfrak{su}(N)$ and $r_i$ are the ranks of the $\mathfrak{su}(r_i)$ gauge nodes of the six-dimensional linear quiver in the tensor branch, the $a_\Sigma$ anomaly we just obtained in \eqref{eq:aS-inflow} matches the gravity computation \eqref{eq:gravres}.

\section{Discussion}

In this work, we have characterized the probe limit of large $\CN=(0,4)$ two-dimensional superconformal defects in six-dimensional $\CN=(0,1)$ SCFTs at large $N$. We have constructed BPS solutions for embedding AdS$_3$ probe D4-branes in AdS$_7\times M_3$ solutions of type IIA SUGRA with non-trivial Romans mass \cite{Apruzzi:2013yva}.  For these probe brane embeddings, we have computed the probe brane contribution to the holographic EE of a spherical region in the dual field theory, the probe brane on-shell action for various AdS$_3$ boundary geometries, and the two-dimensional gravitational anomaly coming from the probe WZ action.  The defect EE and on-shell action uniquely fix the independent defect Weyl anomalies $a_\Sigma$ and $d_2$ in \eq{gravres} at large $N$, while the gravitational anomaly provides a subleading correction to $a_\Sigma$ in \eq{a-sigma-grav}.  These results provide novel holographic predictions for universal quantities that are crucial for the study of the ubiquitous string-like defects in six-dimensional SCFTs. 

With our holographic predictions for defect anomalies in hand, \sn{ft-anomalies} focused on computations in field theory to verify our results by utilizing the various string theoretic construction at our disposal.  In particular on the field theory side, we found that the expectation value of a circular Wilson loop in $\CN=1$ SYM theory on $\mathds{S}^5$ computed using supersymmetric localization agreed exactly with the holographic computation of the on-shell action of a probe D4-brane wrapping $\mathds{S}^1\times\mathds{S}^1$ at the conformal boundary of AdS$_3$. This match mirrors similar computations done for Wilson surface operators in $A_{N-1}$ six-dimensional $\CN=(0,2)$ SCFTs, e.g \cite{Mori:2014tca}.  In string theoretic constructions, we were able to compute defect 't Hooft anomalies using inflow arguments. Working in IIB, we showed that D3-branes wrapping non-compact cycles receive a contribution from inflow to the R-anomaly in their defect four-form anomaly polynomial that, using the non-perturbative relations between defect 't Hooft and Weyl anomalies in \cite{Wang:2020xkc}, matches the leading large $N$ results for $a_\Sigma$ obtained using probe brane holography. 

The inflow contribution provides the leading order contribution of $a_{\Sigma}$ at large $N$ and $r_i$ and we argue that inflow captures the bulk-defects modes. On the other hand the subleading contribution coming from the Wess--Zumino action of the brane is reproduced by the gravitational anomaly of the 2d quiver description of the system with the defect inserted, which has been derived from the NS5--D2--D4--D6--D8 brane system. The 2d quiver gravitational anomaly does not capture the inflow term and the anomaly inflow does not capture the gravitational anomaly contribution to $a_{\Sigma}$. For this reason, we argue that the 2d quiver describes the intrinsic modes of the defect and the anomaly inflow the bulk-defect degrees of freedom. The latter are leading whereas the former are subleading at large $N,r_i$.

\section*{Acknowledgements}

We thank J.~Heckman and H.~Shimizu for their initial involvement in this paper, which is seeing the light after an unusually long gestation.  We express our thanks to Y.~Tachikawa for useful comments on the draft.
This work is supported in part by the INFN. The work of FA is supported in part by the Italian MUR Departments of Excellence grant 2023-2027 "Quantum Frontiers”. NM is partially supported by the MUR-PRIN grant No. 2022NY2MXY. AT is supported in part by MUR-PRIN contract 2022YZ5BA2.

\appendix

\section{Coordinate Transformations}\label{app:conformal}

Let's consider the flat Euclidean geometry
\begin{align}\label{eq:flat-metric}
    ds^2 = \delta_{\mu\nu}dx^\mu dx^\nu
\end{align}
where $\mu,\nu = 1,\ldots, 6$. We start from a flat defect in the $x^5$--$x^6$ plane situated at $x^1 =x^2 = x^3 = x^4 =0$.  Consider the conformal transformation $x^\mu \to \tilde{x}^\mu(x)$ that leaves $\mathbb{R}^6$ invariant but maps the defect submanifold from $\mathbb{R}^2 \to\mathds{S}^2$ with radius $R$:
\begin{align}
\begin{split}
    \tilde x^1  &= R \frac{4 |x|^2 -R^2}{R^2 +4|x|^2 +4 R x^1}\,,\qquad\qquad
    \tilde{x}^a = 4R^2 \frac{x^a}{R^2 +4|x|^2 +4 R x^1}\,,  
\end{split}
\end{align}
where $a = 2,\ldots, 6$ with inverse transformation 
\begin{align}
    \begin{split}
        x^1 &= R \frac{R^2 - |\tilde{x}|^2}{2((R-\tilde{x}^1)^2 +\delta_{bc}\tilde{x}^b\tilde{x}^c)}\,,\qquad\qquad
        x^a= R^2 \frac{\tilde{x}^a}{(R-\tilde{x}^1)^2 +\delta_{bc}\tilde{x}^b\tilde{x}^c}\,.
    \end{split}
\end{align}
The effect of this transformation is to map \eq{flat-metric} to
\begin{align}\label{eq:conformal-S2-R6-metric}
    ds^2 =\frac{R^4}{((R-\tilde{x}^1)^2 + \delta_{bc}\tilde{x}^b \tilde{x}^c)^2}\delta_{\mu\nu}d\tilde{x}^\mu d\tilde{x}^\nu\,.
\end{align}
Importantly the hyperplane $x^1=\ldots =x^4 =0$ is mapped to $\tilde{x}^2 = \tilde{x}^3 = \tilde{x}^4 = 0$ with $\delta_{\mu\nu}\tilde{x}^\mu \tilde{x}^\nu = R^2$. 

To make the defect geometry a bit clearer, use $(\tilde{x}^1)^2 = R^2 - (\tilde{x}^5)^2-(\tilde{x}^6)^2$ and $\tilde{x}^2=\tilde{x}^3 = \tilde{x}^4=0$ to write the image of the line element of the defect submanifold geometry under the conformal map as
\begin{align}
    ds^2|_\Sigma = R^2\frac{(R^2 - (\tilde{x}^6)^2)(d\tilde{x}^5)^2+(R^2 - (\tilde{x}^5)^2)(d\tilde{x}^6)^2+2\tilde{x}^5\tilde{x}^6d\tilde{x}^5 d\tilde{x}^6}{4(R^2 - (\tilde{x}^5)^2-(\tilde{x}^6)^2)\Big(R-\sqrt{R^2 - (\tilde{x}^5)^2-(\tilde{x}^6)^2}\Big)^2}\,.
\end{align}
Then, using the coordinate transformation 
\begin{align}
    \tilde{x}^5 = R \sin\Theta\sin\Phi,\qquad \tilde{x}^6 = R\sin\Theta \cos\Phi
\end{align}
we find
\begin{align}
    ds^2|_{\Sigma} = \frac{R^2}{16 \sin^4\Theta/2}(d\Theta^2 + \sin^2\Theta d\Phi^2)\,,
\end{align}
where the conformal factor can be removed by a residual ${\rm SO}(2,2)$ transformation. So, the defect submanifold embedding now takes the form $\mathds{S}^2 \hookrightarrow \mathbb{R}^6$.  

\begin{figure}[t]
\centering	
\includegraphics[width=0.7\textwidth]{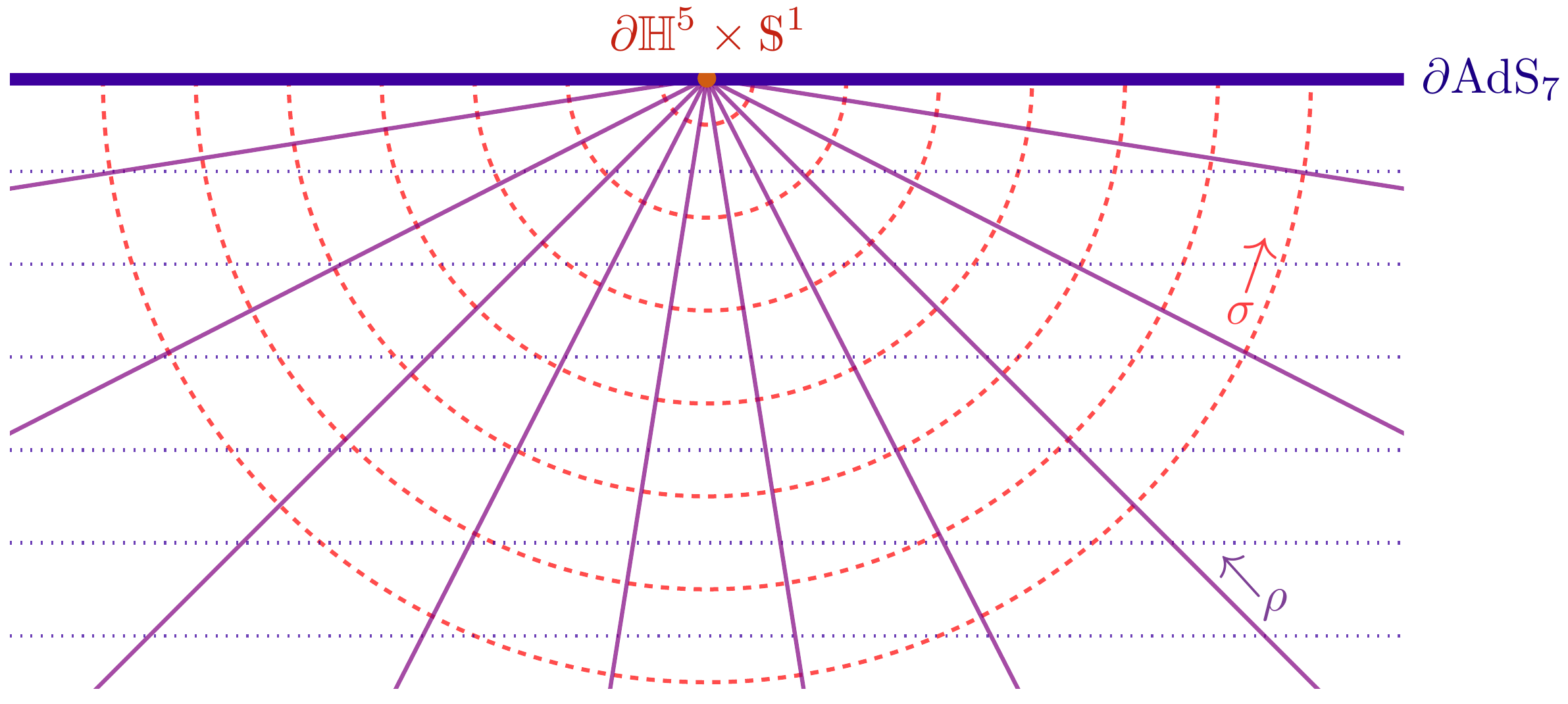}
\caption{\small Starting from the flat $\mathbb{R}^6$ slicing (horizontal blue dotted lines) of AdS$_7$, the map in eqs.~\ref{eq:R6-to-H5-S1-1} and \ref{eq:R6-to-H5-S1-2} brings us to $\mathds{H}_5\times\mathds{S}^1$ sliced AdS$_7$.  The conformal boundary $\partial\text{AdS}_7$ in this new coordinatization can be reached by taking either $\sigma\to\infty$ at fixed $\rho$ which lands in the bulk of $\mathds{H}_5$ or $\rho\to\infty$ at fixed $\sigma$ which simultaneously takes us to $\partial\mathds{H}_5\times\mathds{S}^1$. }
\label{fig:conf-trafo-ads}
\end{figure}

Since we are ultimately interested in studying the geometry of the probe brane dual to this defect, we write the AdS$_7$ metric as \eqref{eq:AdS7-metric-H5-S1}, with
\begin{align}
    ds_{\mathds{H}^5}^2 &= d\rho^2 + \sinh^2\rho d\Omega_4^2\,,\nonumber \\
    \label{eq:AdS7-metric-H5-S1-unpacked}
    d\Omega_4^2 &= d\zeta^2 +\sin^2\zeta d\tilde\Omega_2^2 +\cos^2\zeta d\chi^2\,,\\
    \nonumber
    d\tilde\Omega_2^2 &= d\tilde{\theta}^2 + \sin^2\tilde\theta d\tilde{\phi}^2\,.
\end{align}
 Starting from \eq{flat-metric}, we first map to polar coordinates with
\begin{align}\label{eq:R6-to-H5-S1-1}
    \begin{split}
        &x^1 = \varrho \cos\zeta \cos\chi\,,\quad x^2 = \varrho\cos\zeta\sin\chi\,,\\
        & x^3 = \varrho\sin\zeta \sin\tilde\theta\sin\tilde\phi\,, \quad x^4 = \varrho\sin\zeta\sin\tilde\theta\cos\tilde\phi\,,\quad x^5 = \varrho\sin\zeta \cos\tilde\theta\,,
    \end{split}
\end{align}
which brings us to
\begin{align}\label{eq:transformed-ds6}
    ds_6^2 = d\varrho^2 +\varrho^2d\Omega_4^2 + (dx^6)^2\,.
\end{align}
Now, we can map to $\mathds{H}^5\times \mathds{S}^1$ by
\begin{align}\label{eq:R6-to-H5-S1-2}
   \varrho =   \frac{L\sinh\rho}{\cosh\rho -\cos\varphi}, \qquad x^6 =  \frac{L\sin\varphi}{\cosh\rho -\cos\varphi}\,,
\end{align}
which gives
\begin{align}
    ds_6^2 = \frac{L^2}{(\cosh\rho-\cos\varphi)^2}(d\rho^2 + \sinh^2\rho\, d\Omega_4^2 + d\varphi^2)\,.
\end{align}
Stripping off the overall conformal factor, we can extend this into the bulk with a formula similar to \eqref{eq:CHM-map}:
\begin{equation}
\begin{split}
    \varrho= &\frac{\cosh \sigma \sinh \rho}{\cosh \sigma \cosh\rho+\sinh \sigma \cos\varphi}\,,
    \qquad
    x^6 = \frac{\sinh r \sin\varphi}{\cosh \sigma \cosh\rho+\sinh \sigma \cos\varphi}\,,\\
    z= &\frac{1}{\cosh \sigma \cosh\rho+\sinh \sigma \cos\varphi}\,.
\end{split}
\end{equation}
This turns $(d z^2 + d\varrho^2 +\varrho^2 d\Omega_4^2+dx_6^2)/z^2 $ into \eq{AdS7-metric-H5-S1}. Fig.~\ref{fig:conf-trafo-ads} illustrates this coordinate change. The defect wraps the $\mathds{S}^2\hookrightarrow\mathds{H}_5$ at the conformal boundary of $\mathds{H}_5$ at $\rho\to\infty$, $\zeta = \pi/2$.  The probe brane wraps $\{\rho,\tilde{\theta}, \tilde{\phi}\}$ in the bulk AdS$_7$ geometry and an $\mathds{S}^2$ in the internal space.

We can follow a similar map that brought us to a spherical defect in $\mathds{H}_5\times \mathds{S}^1$ to bring us  to a defect wrapping $\mathds{S}^1\times \mathds{S}^1$ in $\mathds{S}^5 \times \mathds{S}^1$.  We start with \eq{transformed-ds6}, and instead of \eq{R6-to-H5-S1-2}, we take
\begin{align}
   \begin{split}
        \varrho=  \frac{L \sin \varsigma}{\cosh\tau -\cos\varsigma},\qquad x^6 =  \frac{L \sinh \tau}{\cosh\tau - \cos\varsigma}\,,
    \end{split}
\end{align}
which brings us to
\begin{align}
    ds_6^2 = \frac{L^2}{(\cos\varsigma -\cosh\tau)^2}(d\tau^2 + d\varsigma^2 + \sin^2\varsigma d\Omega_4^2)\,.
\end{align}
The boundary metric, up to a conformal factor, is now $\mathbb{R}\times\mathds{S}^5$. Compactifying the $\tau$ direction with periodicity $\tau \sim \tau + \beta$ gives us $\mathds{S}^1_\beta \times \mathds{S}^5$. Extending this geometry into the AdS$_7$ bulk we get
\begin{align}\label{eq:ads7-s1-s5-metric}
    ds_7^2 = d\sigma^2 +\cosh^2\sigma \,d\tau^2 + \sinh^2\sigma\, d\Omega_5^2\,.
\end{align}
The probe now wraps $(\sigma,\,\tau)$ and a great circle at the $\mathds{S}^5$ equator as well as the internal $\mathds{S}^2$.

\section{Two-dimensional quiver gauge theories\label{app:2dquiver}}

As reviewed in Section \ref{sec:ads7},
the six-dimensional theory is realized on the stacks of D6 branes ending on the NS5 branes. The defects we are considering are as in the upper part of Fig.~\ref{fig:brane-diagram-D4D2}. For completeness we also include the string-like objects realized by D2 branes ending on D4 branes.   All in all we end up with the brane intersection in (\ref{tab:string-chain-branes}). 
For simplicity of the analysis, we do not consider the contribution of the D8 branes in the above configuration, but will briefly comment  about their contributions to various anomaly coefficient at the end of this Appendix.

\bes{
\begin{tikzpicture}[baseline,font=\footnotesize]
\draw[solid,black] (-6,0.1) to (6,0.1);
\draw[solid,black] (-6,0) to node[xshift = -6.5cm]{D6} (6,0);
\draw[solid,black] (-6,-0.1) to (6,-0.1);
\draw[blue,thick] (-4,-0.6) to (-4, 0.6);
\draw[blue,thick] (-2,-0.6) to (-2, 0.6);
\node at (1,-0.5) {\blue \Large $\cdots$};
\draw[blue,thick] (2,-0.6) to (2, 0.6);
\draw[blue,thick] (4,-0.6) to (4, 0.6);
\draw[blue,thick] (0,-0.6) to node[below,yshift=-0.6cm]{NS5} (0, 0.6);
\node at (-5, -0.45) {$r_0$};
\node at (-3, -0.45) {$r_1$};
\node at (-1, -0.45) {$r_2$};
\node at (3, -0.45) {$r_N$};
\node at (5, -0.45) {$r_{N+1}$};
\draw[very thick,solid,red] (-6,0.28) to node[xshift = -6.5cm]{D2} (6,0.28);
\node at (-5, 0.5) {\red $Q_0$};
\node at (-3, 0.5) {\red $Q_1$};
\node at (-1, 0.5) {\red $Q_2$};
\node at (3, 0.5) {\red $Q_N$};
\node at (5, 0.5) {\red $Q_{N+1}$};
\draw[snake it, darkgreen,thick] (-2.5,-0.6) to node[above, yshift=0.6cm]{$q_1$} node[below, yshift=-0.6cm]{D4} (-2.5, 0.6);
\draw[snake it, darkgreen,thick] (-0.5,-0.6) to node[above, yshift=0.6cm]{$q_2$} (-0.5, 0.6);
\draw[snake it, darkgreen,thick] (3.5,-0.6) to node[above, yshift=0.6cm]{$q_N$} (3.5, 0.6);
\end{tikzpicture}
}
\begin{equation} \label{tab:string-chain-branes}
\centering
\begin{tabular}{|c||c|c|c|c|c|c|c|c|c|c|}\hline
       &0&1&2&3&4&5&6&7&8&9\\\hline
     D6& $\times$ & $\times$& $\times$& $\times$& $\times$& $\times$& $\times$& -& -& -  \\\hline
     D2& $\times$& $\times$& -& -& -& -& $\times$& -& -& - \\\hline
     D4& $\times$& $\times$& -& -& -& -& -& $\times$& $\times$& $\times$ \\\hline
     NS5& $\times$& $\times$& $\times$& $\times$& $\times$& $\times$& -& -& -& - \\\hline
\end{tabular}
\end{equation}

The two-dimensional $\CN=(0,4)$ gauge theory on the D2 brane worldvolume is described by the linear quiver in (\ref{fig:string-chain-quiver}), which can be obtained in a similar way to those in \cite{Tong:2014yna} and \cite{Haghighat:2013gba}.
\bes{ \label{fig:string-chain-quiver}
\includegraphics[width = 0.75\textwidth]{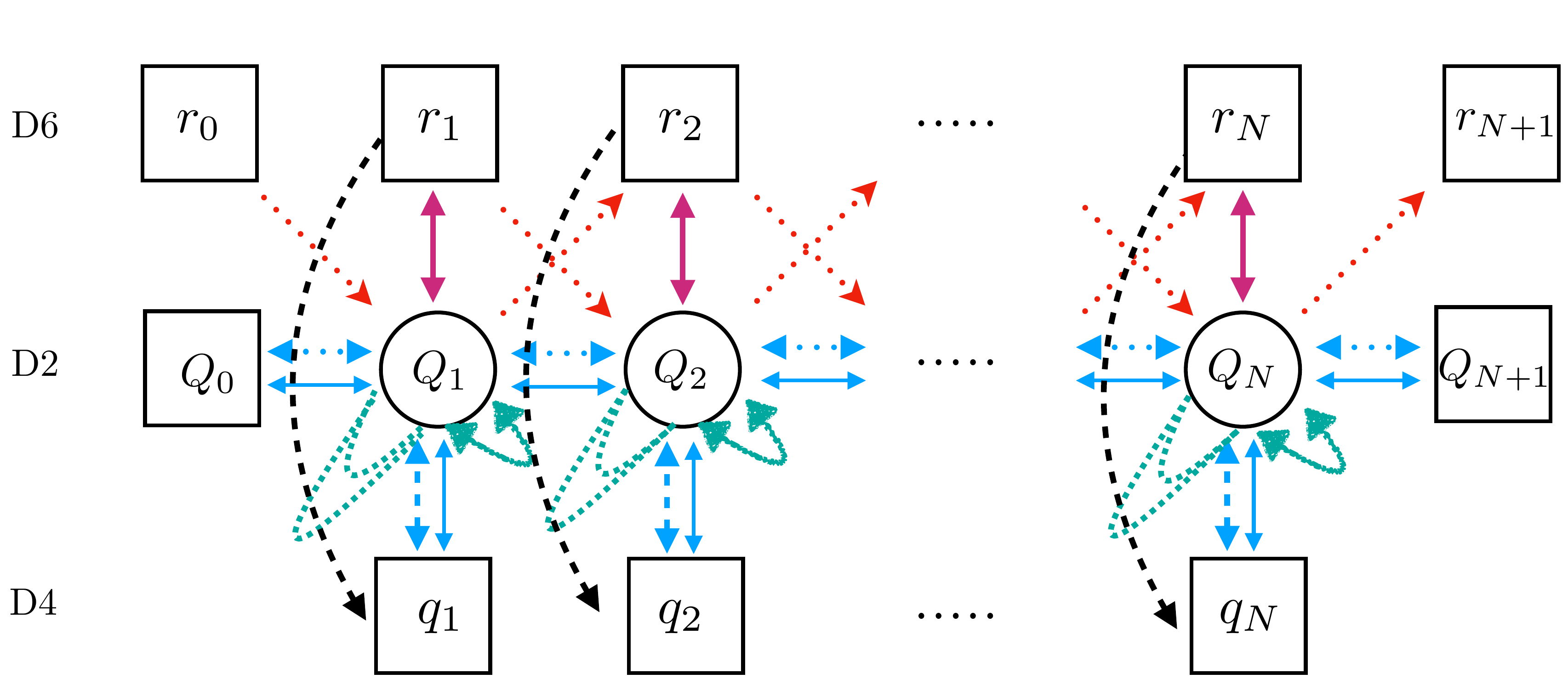}
}

Let us explain the above quiver diagram as follows. First, we have the circular (gauge) nodes in the center row, which are stacks of $Q_i$  D2 branes (the $Q_0$ and $Q_{N+1}$ nodes give rise to $U(Q_0)$ and $U(Q_{N+1})$ flavor symmetries respectively).  The  boxed nodes on the top row are stacks of $r_i$ D6-branes.  The  boxed nodes in the bottom row are stacks of $q_i$ D4-branes.  The length of the row is determined by the number of NS5 branes ($N$). We take $0\leq i \leq N+1$ with $q_0 = q_{N+1} =0$.  We label each field in the quiver as follows.
\begin{equation} \label{fieldlabels}
\includegraphics[width=0.45\textwidth]{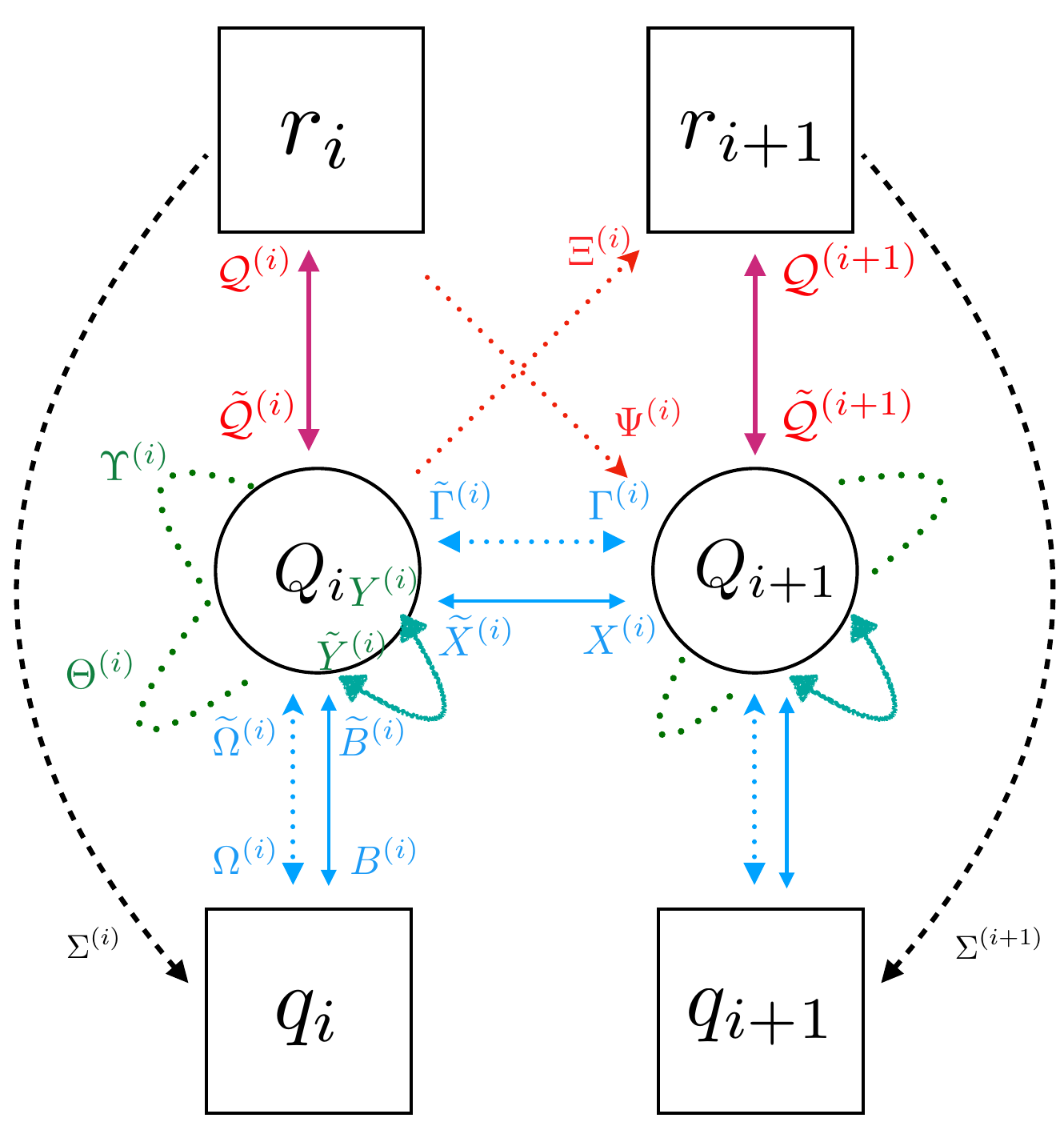}
\end{equation}
\begin{itemize}
\item The $\CN = (0,4)$ vector multiplet consists an $\CN = (0,2)$ vector multiplet $U^{(i)}$ and an adjoint $\CN = (0, 2)$ Fermi multiplet $\Theta^{(i)}$. The vector multiplet $U^{(i)}$ also contains the $\CN=(0,2)$ field strength Fermi multiplet $\Upsilon^{(i)}$. These Fermi multiplets are denoted by dashed green loops in the above diagram.

\item The solid green lines are $\CN=(0,4)$ twisted hypermultiplets built out of $\CN=(0,2)$ chirals $(Y^{(i)},\,\tilde{Y}^{(i)})$ transforming in the adjoint representation of $U(Q_i)$. 

\item The horizontal lines connecting the $Q_i$ nodes comprise an $\CN=(0,4)$ hypermultiplet in the bifundamental of $U(Q_i)\times U(Q_{i+1})$ which in $\CN=(0,2)$ language contains chirals $(X^{(i)}, \tilde{X}^{(i)})$ [solid] and Fermi multiplets $(\Gamma^{(i)}, \tilde\Gamma^{(i)})$ [dashed].
\item The red dashed diagonal lines from the top left corner to bottom right corner are Fermi multiplets $\Psi_i$ charged under $U(r_{i-1})\times U(Q_i)$.  The red dashed diagonal lines from the bottom left corner to top right corner are Fermi multiplets $\Xi_i$ charged under $U(Q_i)\times U(r_{i+1})$.
\item The vertical red solid lines are $\CN=(0,4)$ hypermultiplets in the bifundamental of $U(r_i)\times U(Q_i)$ built out of chiral multiplets $(\CQ^{(i)},\tilde{\CQ}^{(i)})$.
\item The vertical black dashed lines are Fermi multiplets $\Sigma_i$ charged under $U(r_i)\times U(q_i)$.  If there are no D4 branes in the game $q_i=0$ for all $i$, then these multiplets are absent.
\item The vertical blue lines are $\CN=(0,4)$ hypermultiplets charged in the bifundamental of $U(q_i)\times U(Q_i)$.  These multiplets decompose in $\CN=(0,2)$ language as a Fermi multiplet $(\Omega^{(i)},\,\tilde{\Omega}^{(i)})$ and chiral multiplet $(B^{(i)},\tilde{B}^{(i)})$.
\end{itemize}

The E-term potentials for the various Fermi multiplets and superpotential terms can be written in a similar way as in \cite{Tong:2014yna}.  Explicitly, the E-terms are
\bes{
\begin{tabular}{ll}
$E_{\Theta^{(i)}} = [Y^{(i)},\tilde{Y}^{(i)}] + \cQ^{(i)} \tilde{\cQ}^{(i)}$ & \quad
 $E_{\Gamma^{(i)}} = Y^{(i)} X^{(i)} - Y^{(i)} X^{(i-1)}$ \\
  $E_{\tilde{\Gamma}{(i)}} = -\tilde{X}^{(i)}Y^{(i)} +\tilde{X}^{(i-1)}Y^{(i)}$ & \quad
${E}_{{\Xi}^{(i)}}  = X^{(i)} \cQ^{(i+1)}$ \\
${E}_{{\Psi}^{(i)}}  =\tilde{\cQ}^{(i)} {X}^{(i)}$ & \quad
 $E_{\Omega^{(i)}} = Y^{(i)} B^{(i)} - Y^{(i)} B^{(i-1)}$ \\
$E_{\tilde{\Omega}{(i)}} = -\tilde{B}^{(i)}Y^{(i)} +\tilde{B}^{(i-1)}Y^{(i)}$ & \quad $E_{\Sigma^{(i)}} = \tilde{\cQ}^{(i)} B^{(i)}$
\end{tabular}
}
and the superpotential terms are
\bes{ \label{supterms}
\scalebox{0.9}{
\begin{tabular}{ll}
${\cal W}_{\Theta^{(i)}} =  \tilde{X}^{(i)} \Theta X^{(i)}-  X^{(i-1)} \Theta  \tilde{X}^{(i-1)} $  
&${\cal W}_{\Gamma^{(i)}} = \tilde{X}^{(i)} \tilde{Y}^{(i)} \Gamma^{(i)} - \tilde{\Gamma}^{(i-1)}  \tilde{Y}^{(i)}  X^{(i-1)}$  \\
${\cal W}_{\tilde{\Gamma}^{(i)}}  =\tilde{\Gamma}^{(i)}\tilde{Y}^{(i)}X^{(i)} - \tilde{X}^{(i-1)} \tilde{Y}^{(i)}{\Gamma}^{(i-1)}$ &
${\cal W}_{{\Xi}^{(i)}}  = \Xi^{(i)} \tilde{\mathcal{Q}}^{(i+1)} \tilde{X}^{(i)}$  \\
${\cal W}_{{\Psi}^{(i)}}  =  \Psi^{(i)} \tilde{X}^{(i)} \mathcal{Q}^{(i)}$ &
${\cal W}_{\Omega^{(i)}} = \tilde{B}^{(i)} \tilde{Y}^{(i)} \Omega^{(i)} - \tilde{\Omega}^{(i-1)}  \tilde{Y}^{(i)}  B^{(i-1)}$ \\
${\cal W}_{\tilde{\Omega}^{(i)}}  =\tilde{\Omega}^{(i)}\tilde{Y}^{(i)}B^{(i)} - \tilde{B}^{(i-1)} \tilde{Y}^{(i)}{\Omega}^{(i-1)}$ 
& ${\cal W}_{\Sigma^{(i)}} = \Sigma^{(i)} \tilde{B}^{(i)} \mathcal{Q}^{(i)}$\,.
\end{tabular}}
}

\subsection*{Gauge Anomaly Cancellation}
Let us compute the anomaly for quiver \eqref{fig:string-chain-quiver}.  First, the gauge anomaly of the $U(1)_i$ subgroup of the $U(Q_i)$ gauge group is given by
\begin{align} \label{eq:ugan}
    \Tr(\hat{\gamma}^3 J_{U(1)_i}J_{U(1)_i}) = \sum_{j: \, \text{chiral}} \mf{q}_j^2 - \sum_{j: \, \text{Fermi}} \mf{q}_j^2 =2r_i - r_{i+1}-r_{i-1}
\end{align}
where $\hat\gamma^3$ is the two-dimensional chirality matrix.  The gauge anomaly cancellation thus determines $2r_i = r_{i+1} + r_{i-1}$ and is independent of $q_i$. 

This last point deserves a bit of comment.  This result is quite different from the gauge anomaly cancellation in \cite{Gadde:2015tra}.  The key difference is that there is an additional pair of chiral multiplets $(B,\tilde B)$ in the construction above that soaks up the anomaly from the $(\Omega, \tilde\Omega)$ Fermi multiplets which are present in \cite{Gadde:2015tra}.  In the latter case, the $n_i$ are fixed by gauge anomaly cancellation to be $2r_i - r_{i+1}-r_{i-1}$.   

The non-abelian gauge anomaly is obviously more involved but we have all the information we need from the list of matter multiplets above:
\begin{align} \label{eq:sugan}
    \begin{split}
      \hspace{-0.5cm}  \Tr(\hat{\gamma}^3 J_{{\rm SU}(Q_i)}J_{{\rm SU}(Q_i)})=&~ (T_{\CQ}(\Box) +T_{\tilde\CQ}(\bar\Box))r_i - T_{\Xi}(\Box)r_{i+1} - T_{\Psi}(\bar\Box)r_{i-1}\\
        &+T_X(\Box)Q_{i-1} +T_{\tilde X}(\bar\Box)Q_{i+1}-T_\Gamma(\Box)Q_{i-1} +T_{\tilde \Gamma}(\bar\Box)Q_{i+1}\\
        &+(T_{B}(\Box)+T_{\tilde B}(\bar\Box))q_i -(T_{\Omega}(\Box)+T_{\tilde \Omega}(\bar\Box))q_i \\
        &+T_Y({\bf{adj}})+T_{\tilde Y}({\bf{adj}})- T_\Upsilon({\bf{adj}}) -T_\Theta({\bf{adj}}) \\
        =&~r_i -\frac{1}{2}(r_{i+1}+r_{i-1}),
    \end{split}
\end{align}
where $T_F(\mathbf{R})$ denotes the index of the representation $\mathbf{R}$ of $\SU(Q_i)$ in which the field $F$ transforms. We used the convention that the indices of the fundamental and antifundamental representations are $T(\Box)= T (\bar \Box) = \frac{1}{2}$ and that of the adjoint representation is $T({\bf adj})=Q_i $.  This reproduces the abelian gauge anomaly cancellation. 

\subsection*{R-symmetry}
The two-dimensional $\CN=(0,4)$ gauge theory in question has R-symmetry $\so(4)_R \cong \su(2)^+_R \times \su(2)^-_R $. Let $R^\pm[F]$ be the charge of the field $F$ under the Cartan subalgebra of $\su(2)^\pm_R$. The $R$-charge assignment to each field is constrained by the following conditions:
\begin{itemize}
\item We assume the $R$-charges of the fields are independent of the superscript $(i)$, and so we write $R^\pm[F^{(i)}]=R^\pm[F]$ for any field $F$.
\item Each superpotential term carries R-charge $R^\pm[{\cal W}]=+1$.
\item For each Fermi multiplet  $f$, the corresponding $E$-term has charge $R^\pm[E_f] = R^\pm[f] + 1$.
\item $R^\pm[\Upsilon] =1$ since $\Upsilon$ is in the multiplet that contains the field strength.
\end{itemize}
As pointed out in \cite{Tong:2014yna}, the fields $\Gamma^{(i)}$ and $\tilde{\Gamma}^{(i)}$ are singlets under $\su(2)^+_R \times \su(2)^-_R$:
\bes{ \label{RGamma}
R^\pm[\Gamma]=0\,, \quad R^\pm[\tilde{\Gamma}]=0\,.
}
Moreover, as pointed out in \cite{Witten:1994tz, Putrov:2015jpa} and \cite[p. 10]{Tong:2014yna}, it is possible to have a single $\CN = (0, 2)$ Fermi multiplet which is consistent with $\CN = (0, 4)$ supersymmetry. However, for this to happen, the chiral fermion should be a singlet under  $\su(2)^+_R \times \su(2)^-_R$. Thus,
\bes{ \label{RXiPsi}
R^\pm[\Xi]=0\,, \qquad R^\pm[\Psi]=0\,.
}

Taking these into account, we obtain the following R-charge assignment:
\bes{ \label{tab:string-chain-R-charge}
\scalebox{0.9}{
    \begin{tabular}{|c||c|c|c|c|c|c|c|c|c|c|c|c|c|c|c|c|c|}\hline
     & $\Upsilon$ & $\Theta$ & $Y$ & $\tilde Y$ & $X$ & $\tilde X$ & $\CQ$ & $\tilde \CQ$ & $B$ & $\tilde B$ & $\Gamma$ & $\tilde \Gamma$ & $\Omega$ & $\tilde \Omega$ & $\Xi$ & $\Psi$& $\Sigma$\\\hline
        $R^+$ & 1 & 1 & 1 & 1 & 0 & 0 & 1 & 1 & 0 &0 & 0 & 0 &0 &0 &0 &0 &0  \\\hline
        $R^-$ & 1 & $-1$ & 0 &0 & 1 & 1 &0 & 0 & 1 & 1 &0 &0 &0 &0 &0 &0  &0\\\hline
\end{tabular}}
}

\subsection*{$\SU(2)_D$ flavor symmetry}
Our quiver theory also has an ${\rm SU}(2)_D$ flavor symmetry that transforms the $\CN = (0, 4)$ adjoint twisted  hypermultiplet $(Y^{(i)}, \tilde{Y}^{(i)})$ as a doublet for each $i$. We assign their charges under the $\SU(2)_D$ Cartan subalgebra as
\bes{
D[Y^{(i)}] = 1\,, \qquad D[\tilde{Y}^{(i)}] = -1\,.
}
Since the superpotential terms have transform trivially under this symmetry, from \eqref{supterms} we find that the consistent charge assignment is as follows:
\bes{ \label{tab:string-chain-D-charge}
\scalebox{0.9}{
    \begin{tabular}{|c||c|c|c|c|c|c|c|c|c|c|c|c|c|c|c|c|c|}\hline
     & $\Upsilon$ & $\Theta$ & $Y$ & $\tilde Y$ & $X$ & $\tilde X$ & $\CQ$ & $\tilde \CQ$ & $B$ & $\tilde B$ & $\Gamma$ & $\tilde \Gamma$ & $\Omega$ & $\tilde \Omega$ & $\Xi$ & $\Psi$& $\Sigma$\\\hline
        $D$ & 0 & 0 & 1 & $-1$ & 0 & 0 & 0 & 0 & 0 & 0 & 1 & 1 &1 &1 &0 &0 &0  \\\hline
\end{tabular}}
}
Note that $\Upsilon^{(i)}$ and $\Theta^{(i)}$, residing in the $\CN=(0,4)$ vector multiplet, transform trivially in the ${\rm SU}(2)_D$ flavor symmetry.

\subsection*{Gravitational, R-symmetry and ${\rm SU}(2)_D$ Anomalies}
The four-form anomaly polynomial of the two-dimensional quiver gauge theory contains the following terms:
\bes{ \label{termsI4}
\frac{1}{2}K_{R^+} c_2(R^+) + \frac{1}{2} K_{R^-} c_2(R^-) + \frac{1}{2} K_{D} c_2(D) + k_{g} p_1(T)\,,
}
where $K_{R^\pm}$ are R-symmetry anomalies, $K_{D}$ is the ${\rm SU}(2)_D$ symmetry anomaly, and $k_g$ is the gravitational anomaly of the quiver theory:
\bes{
\begin{array}{lll}
    K_{R^\pm} = \Tr(\hat\gamma^3 J_{{\rm SU}(2)^\pm_R} J_{{\rm SU}(2)^\pm_R})\,,& \quad K_{D} = \Tr(\hat\gamma^3 J_{{\rm SU}(2)_D} J_{{\rm SU}(2)_D})\,,&\quad k_g = \Tr\, \hat\gamma^3\, \\
    c_2(R^\pm)= c_2(\SU(2)^\pm_R)~, &  \quad
    c_2(D) = c_2(\SU(2)_D)~, & 
\end{array}
}

For the R-anomalies, the left moving fermions in the Fermi multiplets $\psi$ contribute $-R[\psi]^2$ and the fermions in the chiral multiplet $\chi$ contribute $(R[\chi]-1)^2$, where $R[X]=R_X$ denotes either $R^+[X]$ or $R^-[X]$.  The counting results in 
\bes{
\scalebox{0.9}{$
    \begin{split}\label{eq:krr-corrected}
        K_{R}  &= \sum_{i=0}^{N+1}\left[Q_iQ_{i+1}\left((R_X-1)^2+(R_{\tilde X} -1)^2 -R_\Gamma^2 -R_{\tilde\Gamma}^2\right) -Q_ir_{i+1}R_{\Xi}^2 -Q_ir_{i-1}R_\Psi^2 \right]\\
      &+\sum_{i =1}^N q_i\left[ Q_i\big((R_B-1)^2+(R_{\tilde{B}}-1)^2 -R_{\Omega}^2 - R_{\tilde\Omega}^2\big) -r_iR_{\Sigma}^2\right]\\
      &+\sum_{i=1}^NQ_i\left[Q_i((R_Y-1)^2 +(R_{\tilde Y}-1)^2 - R_\Upsilon^2 -R_\Theta^2 ) +r_i(R_\CQ-1)^2 +r_i(R_{\tilde\CQ}-1)^2\right]\,.
    \end{split}$}
}
Explicitly, using \eqref{tab:string-chain-R-charge}, we have
\bes{ \label{eq:kRR}
K_{R} = \begin{cases}
2\sum_{i=1}^N r_i Q_i& \quad \text{if $R= R^-$} \\
2\sum_{i=1}^N (q_i -Q_i)Q_i+2\sum_{i=0}^{N}Q_iQ_{i+1}& \quad \text{if $R=R^+$}\,. 
\end{cases}
}

On the other hand, the gravitational anomaly can be obtained from by replacing $(R_F-1)^2$ and $R_F^2$ in \eqref{eq:krr-corrected} by $1$ for every field $F$:
\bes{ \label{eq:kg}
    k_g = - \sum_{i=1}^Nr_i q_i+ \sum_{i=1}^N (2r_i-r_{i+1}-r_{i-1}) Q_i = - \sum_{i=1}^Nr_i q_i\,,
}
where we have used the gauge anomaly cancellation to obtain the last equality.

Finally, the ${\rm SU}(2)_D$ anomaly can be computed by replacing $(R_F-1)^2$ and $R_F^2$ in \eqref{eq:krr-corrected} by $D[F]^2$, where $D[F]$ is the charge of $F$ under the Cartan subalgebra of ${\rm SU}(2)_D$ given by \eqref{eq:krr-corrected} for every field $F$. The result is
\bes{ \label{eq:kDD}
K_{D} = - K_{R^+} = 2\sum_{i=1}^N (Q_i -q_i)Q_i-2\sum_{i=0}^{N}Q_iQ_{i+1} \,.
}

To summarize, the terms \eqref{termsI4} in the four-form anomaly polynomial of the two-dimensional quiver theory can be written as
\bes{ \label{I4explicit}
&\frac{1}{2} K_{D} \left[ c_2(D) - c_2(R^+) \right] + \frac{1}{2} K_{R^-} c_2(R^-) + k_g p_1(T) \\
&= \Bigg( \sum_{i=1}^N (Q_i -q_i)Q_i-\sum_{i=0}^{N}Q_iQ_{i+1} \Bigg) \left[ c_2(D) - c_2(R^+) \right] \\
& \qquad + \Bigg(\sum_{i=1}^N r_iQ_i\Bigg) \, c_2(R^-) + \Bigg(- \sum_{i=1}^Nr_i q_i \Bigg)\, p_1(T)\,.
}

It is instructive to compare this result to \cite[(2.5)]{Shimizu:2016lbw} in the case of $q_i=0$. According to \cite[Page 7]{Shimizu:2016lbw}, the coefficient of $c_2(R^-)$ in \eqref{I4explicit} coincides with that of $c_2(I)$ in \cite{Shimizu:2016lbw}, and so we identify our ${\rm SU}(2)_R^-$ with their ${\rm SU}(2)_I$. Moreover, the coefficient of $\left[ c_2(D) - c_2(R^+) \right]$ when $q_i=0$ can be written as $\frac{1}{2} \sum_{i,j =1}^N C_{ij} Q_i Q_j$ where $C_{ij}$ is the Cartan matrix of the $A_N$ algebra. This is in agreement with the coefficient of $c_2(L)-c_2(R)$ of \cite[(2.5)]{Shimizu:2016lbw} when their $\eta^{ij}$ is taken to be the Cartan matrix $C_{ij}$. We thus identify our ${\rm SU}(2)_D-{\rm SU}(2)_R^+$ with their ${\rm SU}(2)_L-{\rm SU}(2)_R$. Finally, we see that for $q_i=0$ the gravitation anomaly $k_g$ vanishes, in agreement with the coefficient of $p_1(T)$ and $c_2(L)+c_2(R)$ of \cite[(2.5)]{Shimizu:2016lbw} when $\eta^{ij}$ is taken to be the Cartan matrix $C_{ij}$, whose diagonal elements are precisely $2$. This is also consistent with \eqref{eq:kDD}, namely the coefficient of $c_2(D)+ c_2(R^+)$ in our anomaly polynomial is zero. This leads to the identification of our ${\rm SU}(2)_D+{\rm SU}(2)_R^+$ with their ${\rm SU}(2)_L+{\rm SU}(2)_R$. The R-symmetry of the two-dimensional theory in our notation is ${\rm SU}(2)_R^+ \times {\rm SU}(2)_R^-$ and in their notation is ${\rm SU}(2)_R \times {\rm SU}(2)_I$.  We therefore identify
\bes{ \label{mapss}
{\rm SU}(2)_R^- \equiv {\rm SU}(2)_I\,, \quad {\rm SU}(2)_R^+ \equiv {\rm SU}(2)_R\,, \quad {\rm SU}(2)_D \equiv {\rm SU}(2)_L\,.
}
in our and their notations, respectively.

 We can compute the defect anomaly from the formula (\ref{eq:yifanformula}), namely 
 \bes{
 a_\Sigma = 3 k_{r} -\frac{1}{2} k_g~.  
 }
 Note that the anomaly polynomial \eqref{termsI4} contains the terms
$\frac{1}{2}K_{R^+} c_2(R^+) = -\frac{1}{2}K_{R^+} c_1(r)^2 $ and $\frac{1}{2}K_{R^-} c_2(R^-) = -\frac{4}{2}K_{R^-} c_1(r)^2 = -2K_{R^-} c_1(r)^2$, where we have used the fact that $c_2(R^+) = -c_1(r)^2$ and $c_2(R^-) =c_2(I) = -4 c_1(r)^2$; see \eqref{mapss}, \eqref{eq:yifanformula} and \eqref{c2Iandc1r}. Recalling that we adopt the convention $\frac{1}{2} k_r c_1(r)^2$ in the anomaly polynomial, we thus obtain
\bes{
k_r = 
\begin{cases}
-4 K_{R^-} & \quad \text{for\,\, $\SU(2)_R^-$} \\
-K_{R^+} & \quad \text{for\,\, $\SU(2)_R^+$}~. \\
\end{cases}
}
Applying the formulae \eqref{eq:kRR} and \eqref{eq:kg}, we obtain
\bes{
 a_\Sigma = 
\begin{cases} 
 -24\sum_{i=1}^N r_iQ_i+\frac{1}{2}\sum_{i=1}^Nr_i q_i  & \quad \text{if $R=R^-$} \\ 
-6\Big(\sum_{i=1}^N (q_i -Q_i)Q_i+\sum_{i=0}^N Q_i Q_{i+1}\Big) +\frac{1}{2}\sum_{i=1}^Nr_i q_i  & \quad \text{if $R=R^+$}\,.
\end{cases}    
}
We extract the defect anomaly by setting the BPS string charges $Q_i=0$. 

We finally comment on the insertion of $f_i$ D8-branes, which span the $012345\, 789$ directions, in the brane system which engineers 6d theories intersecting the $i$-th D6 segment. The extra contribution to 2d quiver gauge theory is provided by D2--D8 states that are $f_i Q_i$ $(0,2)$ Fermi multiplets.\footnote{The D4--D8 states will not contribute since they are not light 2d degrees of freedom.} In addition, the cancellation of the gauge anomalies \eqref{eq:sugan} and \eqref{eq:ugan} now requires
\begin{equation} \label{eq:gengan}
    2r_i -r_{i+1}-r_{i-1}-f_{i}=0
\end{equation}
which is equivalent to the 6d irreducible  gauge anomaly (proportional to ${\rm tr}(F^4)$ in the 6d anomaly polynomial) cancellation condition.
All the R-charge assignments are the same as $R_{\Xi(i)}$. Therefore,
the $K_{R^{\pm}}$ and $K_{D}$ will receive contributions from $f_i Q_i$ $(0,2)$ Fermi multiplets accordingly. The gravitational anomaly remains unchanged since it explicitly depends on the gauge anomaly cancellation condition \eqref{eq:kg}.
We conclude that the addition of D8 branes does not change the defect anomaly contribution, but it changes the BPS strings R-symmetries anomalies.

\bibliographystyle{JHEP}
\bibliography{6dStrings}
\end{document}